%% file: arxiv.tex
\documentclass[acmtog,screen]{acmart}

\definecolor{myblue}{RGB}{31, 119, 180}
\definecolor{myorange}{RGB}{255, 127, 14}
\definecolor{mygreen}{RGB}{44, 160, 44}
\definecolor{myred}{RGB}{214, 39, 40}

\newcommand{\Rv}[1]{\textcolor[RGB]{0, 0, 0}{{#1}}}

\AtBeginDocument{%
  }

\copyrightyear{2026}
\acmYear{2026}
\setcopyright{cc}
\setcctype{by}
\acmConference[SIGGRAPH Conference Papers '26]{Special Interest Group on Computer Graphics and Interactive Techniques Conference Conference Papers}{July 19--23, 2026}{Los Angeles, CA, USA}
\acmBooktitle{Special Interest Group on Computer Graphics and Interactive Techniques Conference Conference Papers (SIGGRAPH Conference Papers '26), July 19--23, 2026, Los Angeles, CA, USA}
\acmDOI{10.1145/3799902.3811071}
\acmISBN{979-8-4007-2554-8/2026/07}

\acmSubmissionID{371}

\usepackage{prettyref}
\newrefformat{fig}{Figure~\ref{#1}}
\newrefformat{par}{Section~\ref{#1}}
\newrefformat{appen}{Appendix~\ref{#1}}
\newrefformat{sec}{Section~\ref{#1}}
\newrefformat{sub}{Section~\ref{#1}}
\newrefformat{table}{Table~\ref{#1}}
\newrefformat{ass}{Assumption~\ref{#1}}
\newrefformat{alg}{Algorithm~\ref{#1}}
\newrefformat{def}{Definition~\ref{#1}}
\newrefformat{thm}{Theorem~\ref{#1}}
\newrefformat{cor}{Corollary~\ref{#1}}
\newrefformat{lem}{Lemma~\ref{#1}}
\newrefformat{step}{Step~\ref{#1}}
\newrefformat{ln}{Line~\ref{#1}}
\newrefformat{rem}{Remark~\ref{#1}}
\newrefformat{eq}{Equation~\ref{#1}}
\newrefformat{pb}{Problem~\ref{#1}}
\newrefformat{it}{Item~\ref{#1}}
\newrefformat{te}{Term~\ref{#1}}
\def\Eqref Eq:#1:{\eqref{eq:#1}}
\newrefformat{Eq}{Equation~\Eqref#1:}

\newcommand{\xx}{\mathbf{x}}
\newcommand{\oo}{\mathbf{o}}
\newcommand{\zz}{\mathbf{z}}
\renewcommand{\ss}{\mathbf{s}}
\newcommand{\nn}{\mathbf{n}}
\renewcommand{\SS}{\mathbf{S}}
\newcommand{\pp}{\mathbf{p}}

\usepackage[noend]{algpseudocode}
\usepackage{xcolor}
\usepackage{colortbl}
\usepackage[ruled,linesnumbered]{algorithm2e}

\def\BState{\State\hskip-\ALG@thistlm}
\makeatother

\newcommand{\algorithmicforeach}{\textbf{foreach}}
\algdef{SE}[FOREACH]{ForEach}{EndForEach}[1]{\algorithmicforeach\ #1\ \algorithmicdo}{\algorithmicend\ \algorithmicforeach}%
\algtext*{EndForEach}

\definecolor{smGrey}{rgb}{0.8, 0.8, 0.8}  
\definecolor{smOrange}{rgb}{1.0, 0.64, 0.0} 
\definecolor{smBlue}{rgb}{0.0, 0.64, 1.0}  

\begin{document}

\title{HairLRM: Strand-based Hair Modeling via Large Reconstruction Models
}

\author{Yuefan Shen}
\authornote{Both authors contributed equally to this research.} 
\orcid{0000-0002-6049-7966}
\email{yuefanshen@outlook.com}
\affiliation{%
  \institution{LIGHTSPEED}
  \city{Shenzhen}
  \country{China}
}

\author{Yican Dong}
\authornotemark[1] 
\orcid{0009-0005-1433-2553}
\email{22321117@zju.edu.cn}
\affiliation{%
  \institution{State Key Lab of CAD and CG, Zhejiang University}
  \city{Hangzhou}
  \country{China}
}

\author{Xiufeng Huang}
\orcid{0009-0003-2249-3264}
\email{xiufenghuang@life.hkbu.edu.hk}
\affiliation{%
  \institution{Hong Kong Baptist University}
  \city{Hong Kong}
  \country{China}
}

\author{Zhongtian Zheng}
\orcid{0009-0009-4714-1760}
\email{zhengzhongtian@pku.edu.cn}
\affiliation{%
  \institution{LIGHTSPEED}
  \city{Shenzhen}
  \country{China}
}

\author{Youyi Zheng}
\authornote{Corresponding author}
\orcid{0000-0002-9120-9592}
\email{youyizheng@zju.edu.cn}
\affiliation{%
  \institution{State Key Lab of CAD and CG, Zhejiang University}
  \city{Hangzhou}
  \country{China}
}

\author{Kui Wu}
\orcid{0000-0003-3326-7943}
\email{kwwu@lightspeed-studios.com}
\affiliation{%
  \institution{LIGHTSPEED}
  \city{Los Angeles}
  \state{CA}
  \country{USA}
}

\begin{abstract}
The fundamental limitation of traditional strand-based modeling is not simply data scarcity, but the ill-posedness of inferring complex 3D fields from 2D imagery without structural constraints. This unconstrained regression leads to catastrophic failures in resolving both global occlusion (e.g., in ponytails) and local directionality (e.g., in curls), resulting in over-smoothed, plausible-but-incorrect geometries. To resolve this, we integrate the strong geometric priors of Large Reconstruction Models (LRMs) into the strand generation pipeline. Using the LRM mesh as a structural anchor, we employ a novel Dual Orientation AutoEncoder to lift coarse geometry into high-fidelity strands. By resolving vector field singularities through latent-space optimization and surface-guided refinement, our method effectively disentangles complex topological structures, setting a new benchmark for robustness and accuracy in hair reconstruction.
\end{abstract}


\begin{teaserfigure}
\centering
\includegraphics[width=\linewidth]{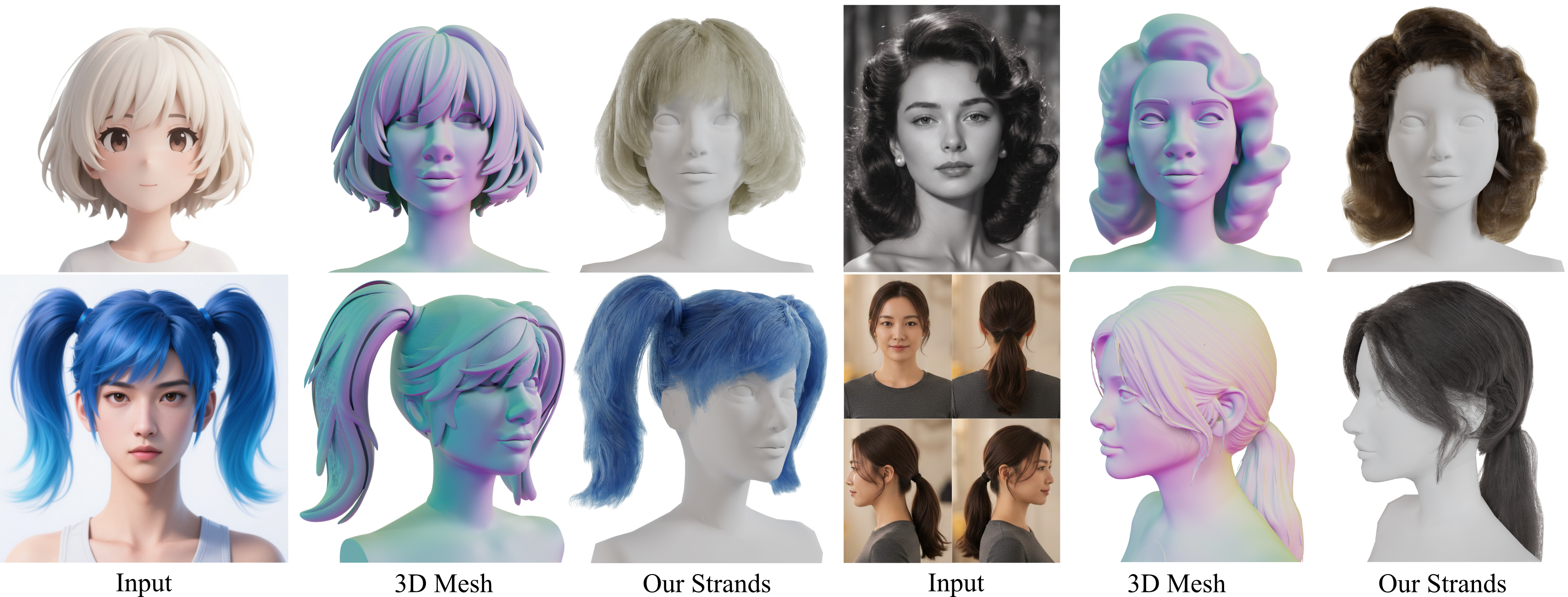}
\caption{Our method reconstructs strand-level hair from image inputs via an intermediate 3D mesh, generalizing effectively across diverse domains (e.g., cartoons, realistic portraits) and complex styles (e.g., curly hair and ponytails). Unlike prior single-view methods, our framework also supports multi-view inputs to ensure geometric consistency.}
\Description{}
\label{fig:teaser}
\end{teaserfigure}

\maketitle
\input{sections/0.intro}

\input{sections/1.bkgd}

\input{sections/2.method}
\input{sections/3.exp_arxiv}

\input{sections/4.conclusion}

\begin{acks}
This work was supported in part by NSF China (No. 62172363). Authors would also like to thank Yujian Zheng for generously sharing the dataset~\cite{zheng2024towards} that facilitated this paper.
The input images were generated via the HunyuanImage 3.0~\cite{cao2025hunyuanimage}.
\end{acks}

\clearpage
\newpage

\begin{figure*}[ht!]
    \centering
    \includegraphics[width=\linewidth]{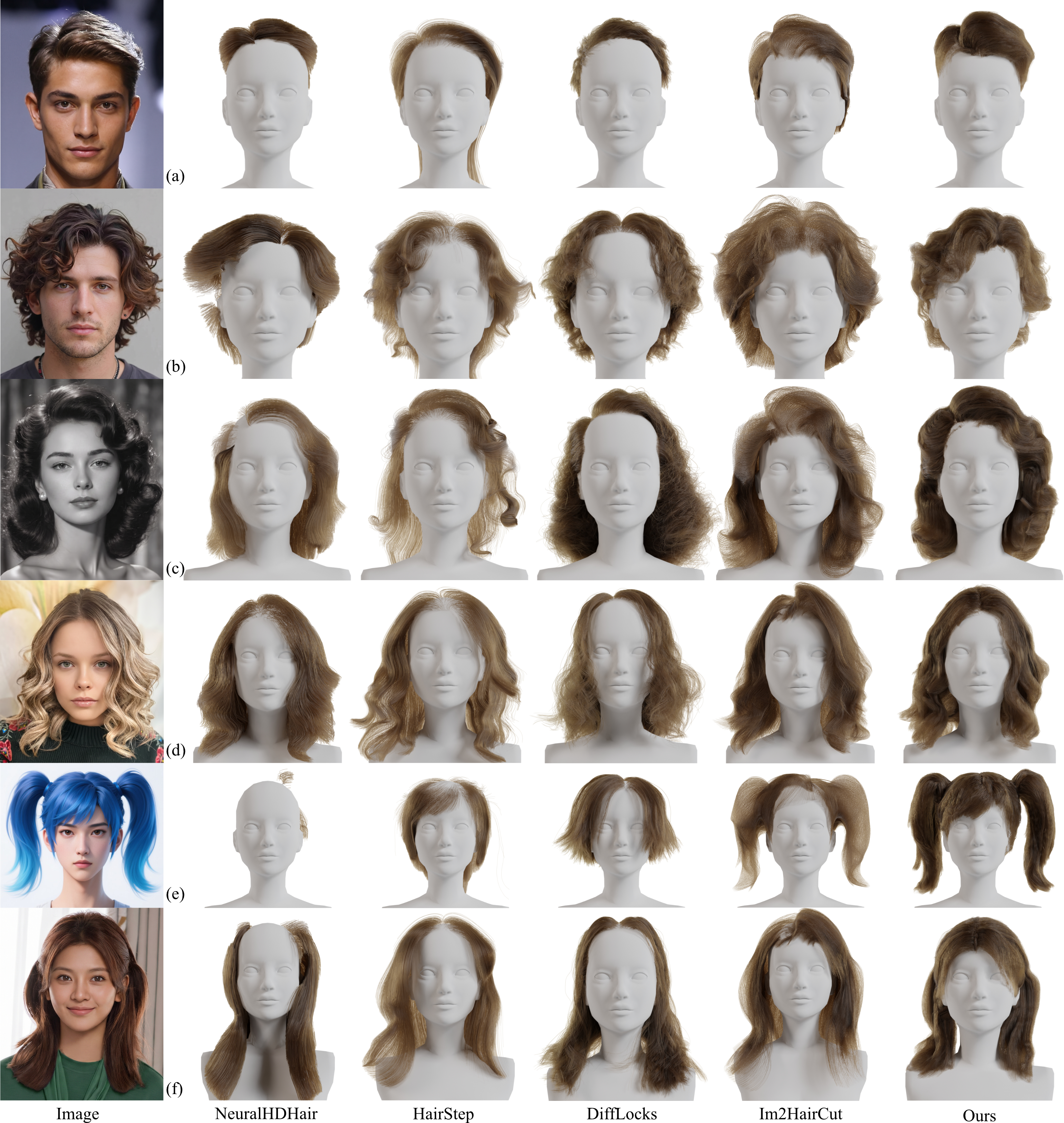}
    \caption{
    Visual comparisons with prior works: NeuralHDHair~\cite{wu2022neuralhdhair}, HairStep~\cite{zheng2023hairstep}, DiffLocks~\cite{rosu2025difflocks} \Rv{and Im2Haircut~\cite{sklyarova2025im2haircut}}.
    }
    \label{fig:Comp}
    \Description{}
\end{figure*}

\clearpage
\newpage

\begin{figure*}[t!]
    \centering
    \includegraphics[width=\linewidth]{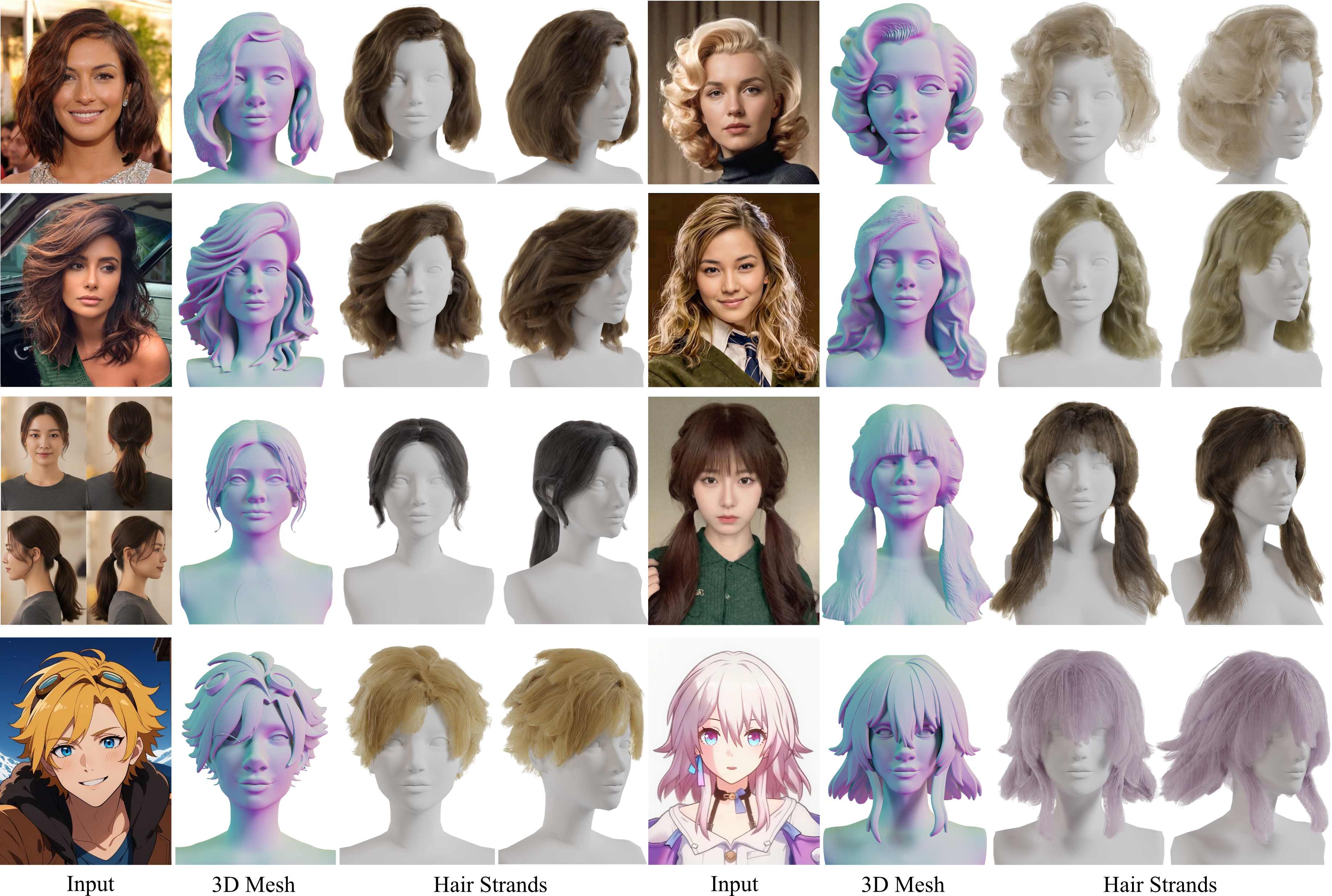}
    \caption{Gallery of reconstructed hairstyles demonstrating the versatility and accuracy of our method across diverse hair types and styles.}
    \label{fig:gallery}
    \Description{}
\end{figure*}

\begin{figure*}[t!]
    \centering
    \includegraphics[width=\linewidth]{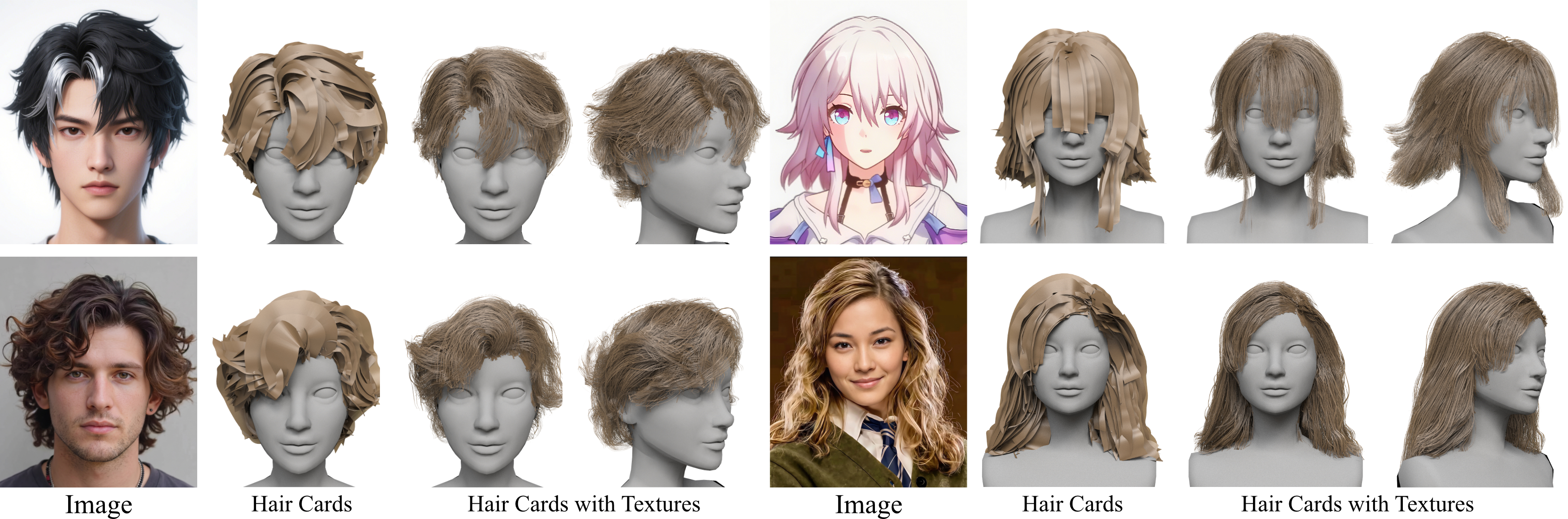}
    \caption{Hair cards converted from our reconstructed strands by~\cite{zheng2025auto}
    }
    \label{fig:cards}
    \Description{}
\end{figure*}

\clearpage
\newpage

\bibliographystyle{ACM-Reference-Format}
\bibliography{bibliography}

\appendix
\input{sections/5.appendix}

\end{document}

%% file: sections/0.intro.tex
\section{Introduction}

The creation of photorealistic digital humans for cinema, interactive entertainment, and virtual presence demands high-fidelity 3D hair modeling. However, achieving this remains a formidable computational challenge, stemming from the intricate topology and sheer scale of human hair, which comprises upwards of 150,000 individual strands exhibiting complex geometric variations, from tight curls to structured arrangements such as ponytails. Consequently, contemporary industrial pipelines remain bottlenecked by labor-intensive manual authoring processes.

While recent data-driven approaches have advanced automatic hair reconstruction from \Rv{monocular~\cite{zhou2018hairnet,zhang2019hair,wu2022neuralhdhair,zheng2023hairstep,sklyarova2025im2haircut} and multi-view inputs~\cite{kuang2022deepmvshair,rosu2022neural,sklyarova2023neural,wu2024monohair,takimoto2024dr,zakharov2024human,zhou2024groomcap,zakharov2024gaussianhaircut}}, fundamental limitations persist. We argue that inaccuracy in current state-of-the-art strand modeling stems not merely from data sparsity, such as reliance on limited synthetic corpora, e.g., the USC Hair Salon dataset with only 343 models~\cite{hu2015single}, but also from the inherently ill-posed nature of regressing high-dimensional 3D vector fields directly from 2D projections without robust structural constraints. This unconstrained formulation results in two critical pathologies: 1) Global Structural Collapse, where missing semantic clues lead to irresolvable ambiguities in depth ordering and occlusion within complex topologies (e.g., braided or ponytail styles); 2) Manifold Ambiguity, where conventional single-vector field representations inherently over-smooth the high-frequency, multi-directional flows characteristic of curly or chaotic hair.
Parallel advancements in Large Reconstruction Models (LRMs) have demonstrated remarkable efficacy in generating high-fidelity 3D surface assets from modest inputs~\cite{chen2024comboverse,lan2024ln3diff,li2024craftsman3d,qu2025stylesculptor,lai2025hunyuan3d}. However, these surface-manifold representations are fundamentally incompatible with the volumetric, strand-based geometry required for realistic hair simulation. Furthermore, deriving accurate strand flows from a coarse manifold surface remains a non-trivial geometric challenge.

To transcend these limitations, we propose a novel framework that bridges this gap by leveraging LRM-generated meshes as an explicit geometric scaffold. We tackle the critical challenge of converting this coarse surface prior into fine-grained strand geometry using a novel Dual Orientation formulation that resolves the directional ambiguities inherent in complex flows. Our architecture centers on a Dual Orientation AutoEncoder (DOAE), trained on a newly synthesized dataset of dual-manifold point clouds, that predicts high-fidelity 3D orientation fields via targeted latent-code optimization. This predicted field is subsequently refined through surface-guided correction and converted into explicit strand geometry. Finally, we introduce a closed-loop iterative refinement strategy that feeds the extracted strands back into the encoder to progressively improve reconstruction accuracy. Extensive validation across diverse input domains—ranging from real-world photography to stylized imagery—demonstrates that our approach achieves superior robustness and fidelity. By effectively disentangling complex geometries, our method significantly outperforms existing state-of-the-art techniques, particularly in capturing accurate global silhouettes and resolving intricate topologies such as ponytails and wavy hair.

%% file: sections/1.bkgd.tex
\section{Related work}

\paragraph{Hair representations}
While mesh-based~\cite{yuksel2009hair} and card-based representations~\cite{zheng2025auto,tojo2025strands2cards} support efficient simulation and rendering~\cite{wu2016real,Bhokare2024hairmesh,huang2024real}, their low-polygon nature inherently limits fine-scale realism. Similarly, recent neural representations~\cite{luo2024gaussianhair,wang2023neuwigs} capture complex geometry but lack the explicit fiber-level structures required for accurate physically based rendering~\cite{marschner2003light} and simulation~\cite{Huang2023,tafuri2019strand,Hsu2023,Hsu2024,Hsu2025}. \Rv{Recent 3D-Gaussian head avatars, such as MeGA~\cite{wang2025mega} and HairCUP~\cite{kim2025haircup}, explore hybrid representations, but cannot be directly consumed by strand-based shading~\cite{marschner2003light} and simulation.} Consequently, strand-based representations remain the standard for high-end production in film and AAA games~\cite{unrealengine}, as they alone provide the geometric fidelity necessary for high-quality shading and dynamic animation.

\paragraph{Hair modeling}
While manual authoring tools like Maya XGen and early heuristic or database-driven methods laid the foundation for hair modeling~\cite{Yuksel2009,fu2007sketching,shen2020deepsketchhair,Kong1998generation,paris2008hair,jakob2009capturing,hu2017avatar,sun2021human,hu2015single,chai2016autohair,liang2018video,Meishvili2024}, the field has largely shifted toward automatic, learning-based reconstruction. Recent deep learning approaches have made significant strides using both single-view~\cite{zhou2018hairnet,zhang2019hair,wu2022neuralhdhair,zheng2023hairstep,sklyarova2025im2haircut} and multi-view inputs~\cite{kuang2022deepmvshair,rosu2022neural,sklyarova2023neural,wu2024monohair,takimoto2024dr,zakharov2024human,zhou2024groomcap,zakharov2024gaussianhaircut}. State-of-the-art methods further leverage latent hair feature representations to enhance detail preservation and diversity~\cite{zhou2023groomgen,sklyarova2024text,chen2024doubly,he2025perm,Coban2025,rosu2025difflocks}. \Rv{Most relevant to ours, Im2Haircut~\cite{sklyarova2025im2haircut} and Gaussian Haircut~\cite{zakharov2024gaussianhaircut} regress strands directly from image observations; we instead condition on an LRM-generated geometric scaffold.} Despite these advances, existing frameworks remain limited: they often fail to capture accurate global structures (e.g., silhouettes) and struggle with complex topologies, leaving hairstyles such as ponytails largely unsupported.
\citet{hu2014} reconstruct braided hairstyles by extracting and matching example patches to recover the hair surface. This work is closest in spirit to ours; however, our approach employs a Dual Orientation Autoencoder to directly predict hair orientations for high-quality strand results.

\begin{figure*}[t!]
\centering
\includegraphics[width=\linewidth]{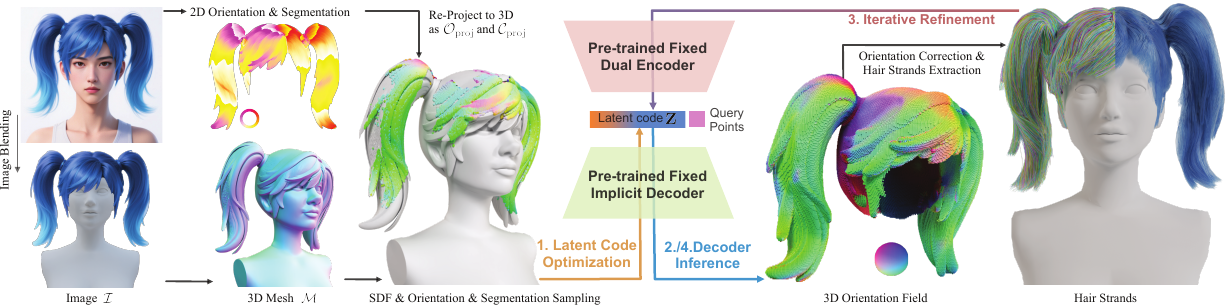}
\caption{The pipeline of our framework. The colorful arrows represent the runtime inference stage of the proposed Dual Orientation AutoEncoder.}
\label{fig:pipeline}
\Description{}
\end{figure*}

\paragraph{Large reconstruction models}
Recent advances in AIGC and 3D representations~\cite{kerbl20233d,mildenhall2021nerf} have enabled large-scale generative models~\cite{hong2023lrm,chen2024comboverse,lan2024ln3diff,li2024craftsman3d,qu2025stylesculptor,lai2025hunyuan3d} capable of producing high-fidelity 3D assets. Many of these approaches~\cite{ren2024xcube,wu2024direct3d,wei2025octgpt} leverage variational autoencoders (VAEs)~\cite{kingma2013auto} to learn compact latent spaces for efficient shape encoding. \Rv{Domain-specialized LRMs such as FaceLift~\cite{lyu2025facelift} can lift a single portrait into a textured multi-view head.} While state-of-the-art methods like Dora-VAE~\cite{chen2025dora} further enhance surface details via dual cross-attention, these mechanisms are explicitly tailored for manifold geometry and do not directly extend to strand-based hair representations. In this paper, we bridge this gap by leveraging the strong semantic and geometric priors learned in these large-scale models. Specifically, we use an intermediate representation, i.e., the output mesh surface, together with our designed dual orientation autoencoder to resolve global structural ambiguities (e.g., depth and occlusion in ponytails) and the directional ambiguity inherent in previous single-vector field approaches, thereby yielding more stable and accurate hair geometry prediction.

%% file: sections/2.method.tex
\section{Method}

An overview of our proposed framework is presented in \autoref{fig:pipeline}. The pipeline begins with an input portrait, from which the hair region is segmented and composited onto a canonical template bust to produce a standardized image, $\mathcal{I}$, adopting the preprocessing protocol established in~\cite{zheng2024towards}. We subsequently leverage a state-of-the-art image-to-3D LRM to derive an initial 3D surface mesh, $\mathcal{M}$, which serves as a coarse geometric prior. \Rv{As LRM naturally accepts multiple views when available, the remainder of the pipeline operates unchanged in both single- and multi-view settings.} The cornerstone of our approach is the novel Dual Orientation AutoEncoder (DOAE), which accepts both $\mathcal{I}$ and $\mathcal{M}$ as inputs. The DOAE is designed to transduce the surface geometry of $\mathcal{M}$ into a high-fidelity volumetric 3D orientation field, $\mathcal{O}$, achieved via latent code optimization (\autoref{sec:DOAE}). Supervision of this architecture is provided by a novel synthetic data generation pipeline that creates dual-manifold hair point clouds (\autoref{sec:data}). Following the prediction of the orientation field $\mathcal{O}$, we extract explicit 3D strand geometry (\autoref{sec:strandExtract}). However, we observe that the initial reconstruction may exhibit pathologies, such as mesh interpenetration and local directional anomalies, attributable to the absence of explicit geometric coupling between the inferred field $\mathcal{O}$ and the surface proxy $\mathcal{M}$. Consequently, we introduce two refinement stages designed to rectify these artifacts, as detailed in~\autoref{sec:correction} and~\autoref{sec:optim}.

\subsection{Dual Orientation AutoEncoder (DOAE)} \label{sec:DOAE}

Our method follows the architecture of Dora~\cite{chen2025dora}, which combines a dual cross-attention mechanism on salient points with the transformer-based 3DShape2VecSet VAE~\cite{zhang20233dshape2vecset} to capture fine-grained geometric details. \Rv{Throughout this paper, the term \emph{dual} refers to the two complementary hair point clouds that jointly condition the encoder: a \emph{uniform} point cloud $\mathcal{P}_\text{uniform}$ that captures the global hair volume, and a \emph{salient} point cloud $\mathcal{P}_\text{salient}$ that emphasises regions with strong directional change (curls, flow boundaries, partings).} Unlike Dora, our goal is to design a network that encodes hair geometry into a compact latent space and decodes it into detailed hair structures conditioned on the surface. Since our input is a reconstructed mesh surface from a portrait image, the sampling domain must cover both the hair volume and the underlying bust. Moreover, because hair is a codimensional structure, tangent information along the strands is essential for representing hair directionality. Accordingly, each sample in our framework is defined as $\ss = [\xx,\oo,c  ]\in \mathbb{R}^7$, comprising 3 dimensions for position $\xx$, 3 for orientation $\oo$, and 1 binary indicator $c$ denoting hair or bust.
Given a 3D surface, our pipeline consists of three key steps:
\paragraph{1. Sampling}
In particular, given a uniform sampled hair point cloud $\mathcal{P}_\text{uniform}$ and its salient hair point cloud $\mathcal{P}_\text{salient}$, we first downsample them separately to a sparse point cloud set $\mathcal{P}_\text{sparse}$ using Farthest Point Sampling (FPS)~\cite{Moenning2003}.

\paragraph{2. Feature encoding} We calculate the cross-attention features between $\mathcal{P}_\text{sparse}$ and the points sampled uniformly and saliently as:
\begin{equation}
\begin{aligned}
    C_\text{uniform} &= \texttt{CrossAttn}(\texttt{PosEmb}(\mathcal{P}_\text{sparse}), \texttt{PosEmb}(\mathcal{P}_\text{uniform})) \;, \\
    C_\text{salient} &= \texttt{CrossAttn}(\texttt{PosEmb}(\mathcal{P}_\text{sparse}), \texttt{PosEmb}(\mathcal{P}_\text{salient})) \;,
\end{aligned}
\end{equation}
where $\texttt{PosEmb}()$ is the position embedding function~\cite{Mildenhall2021}.
Then, we employ self-attention on the combined attention results to calculate the latent code $\zz$:
\begin{equation}
    \Rv{\zz = \texttt{SelfAttn}(C_\text{uniform} + C_\text{salient})} \;.
\end{equation}

\paragraph{3. Geometry decoding} 
Following the classic implicit function formulation \cite{chen2019learning, park2019deepsdf}, but unlike previous works~\cite{zhang20233dshape2vecset, chen2025dora}, which only predict SDF or occupancy values using the implicit decoder, we add orientation and classification predictions to the pipeline. Specifically, given randomly sampled spatial query points $\mathcal{P}_\text{query}$ and use attention layers, our decoder produces:
\begin{equation}
\label{eq:decoder}
\begin{aligned}
    \hat{\mathcal{D}} &= \texttt{CrossAttn}_\mathcal{D}(\texttt{SelfAttn}_\mathcal{D}(\zz), \texttt{PosEmb}(\mathcal{P}_\text{query})) \;, \\
    \hat{\mathcal{O}} &= \texttt{CrossAttn}_\mathcal{O}(\texttt{SelfAttn}_\mathcal{O}(\zz), \texttt{PosEmb}(\mathcal{P}_\text{query})) \;, \\
    \hat{\mathcal{C}} &= \texttt{SoftMax}\bigl(\texttt{CrossAttn}_\mathcal{C}(\texttt{SelfAttn}_\mathcal{C}(\zz), \texttt{PosEmb}(\mathcal{P}_\text{query}))\bigr) \;,
\end{aligned}
\end{equation}
where $\hat{\mathcal{D}}$, $\hat{\mathcal{O}}$, and $\hat{\mathcal{C}}$ are the predicted signed distance, orientation, and classification label on the grid, respectively. Note that $\texttt{CrossAttn}$ and $\texttt{SelfAttn}$ use different network parameters.

\begin{figure*}[t]
\centering
\includegraphics[width=\linewidth]{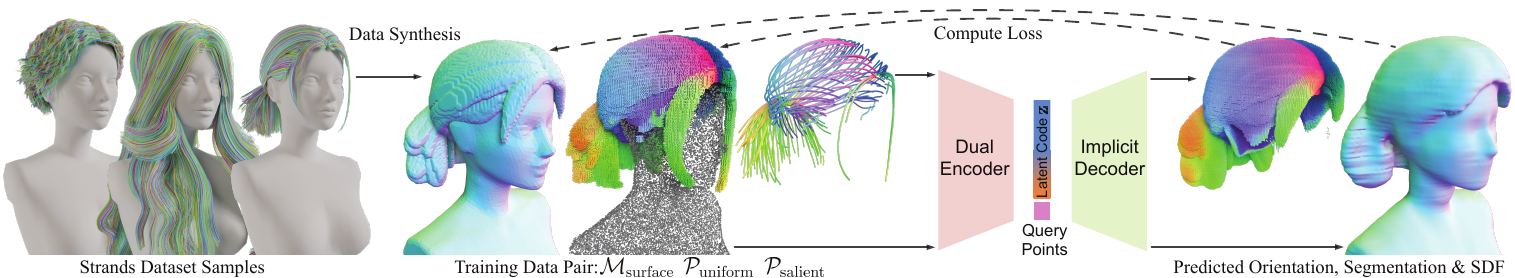}
\caption{We show sampled data, including raw strands data and training data, and illustrate our training pipeline. The sampled strands are from DiffLocks~\cite{rosu2025difflocks}, Perm~\cite{he2025perm}, and UniHair~\cite{zheng2024towards}, respectively. 
}
\label{fig:dataset}
\Description{}
\end{figure*}

\subsection{Data Preparation} \label{sec:data}

To prepare the training data, we collect a total of 52K strand-based hairstyles from three datasets: 40K from DiffLocks~\cite{rosu2025difflocks}, 10K from Perm~\cite{he2025perm}, and 2K from UniHair~\cite{zheng2024towards}. The dataset spans a wide range of styles, including straight, wavy, curly, long, and braided hairstyles. Each hairstyle is first translated and uniformly scaled to approximately align with a given bust mesh $\mathcal{M}_{\text{bust}}$. \autoref{fig:dataset} shows samples of our strands dataset and an overview of our dataset synthesis.
We denote the aligned strand-based hairstyle as $\mathcal{H} = \{ \SS_i \}$, where each strand $\SS_i = \{ \xx_{ij} \}$ consists of sampled points along the strand centerline. The corresponding dense point cloud is denoted as $\mathcal{P}_\text{dense} = \{ \ss_{ij} \}$, where each sample $\ss_{ij} = [\xx_{ij}, \oo_{ij}, 1]$ encodes the position $\xx_{ij}$, the local strand orientation $\oo_{ij}$, and a binary indicator denoting a hair sample.
For training, each hairstyle is further processed to generate a tuple comprising surface mesh $\mathcal{M}_{\text{surface}}$, a salient point cloud $\mathcal{P}_{\text{salient}}$, and a uniformly sampled point cloud $\mathcal{P}_{\text{uniform}}$.

\paragraph{1. Mesh surface $\mathcal{M}_\text{surface}$}
To obtain the corresponding hair mesh, we first compute an occupancy grid from the densified hair strands and extract its iso-surface to form the hair mesh $\mathcal{M}_\text{hair}$. We then perform a boolean union between $\mathcal{M}_\text{hair}$ and the bust mesh $\mathcal{M}_\text{bust}$ to obtain the combined surface $\mathcal{M}_\text{surface} = \mathcal{M}_\text{hair} \bigcup \mathcal{M}_\text{bust}$. This unified mesh is used to compute signed distance field (SDF) values for network training. 

\paragraph{2. Salient point cloud $\mathcal{P}_\text{salient}$}
\citet{chen2025dora} propose a heuristic sharp-edge detection scheme for triangle meshes that samples salient points for high-quality reconstruction using a dual cross-attention mechanism. However, this approach relies on mesh-specific geometric cues and therefore cannot be directly applied to hair data. Inspired by Dora, we introduce a novel salient point sampling strategy tailored to strand-based hair representations. Following \citet{wang2009, zheng2025auto}, we employ a strand-based k-means clustering to partition the input hair into $N_\text{cluster}$ clusters. The distance between two strands $\mathcal{S}_i$ and $\mathcal{S}_j$ is defined as $\gamma(\mathcal{S}_i, \mathcal{S}_j) = \sum_{k=1}^{N^s} \|\xx_{ik} - \xx_{jk}\|_2^2 / N^s$ where $N^s$ denotes the number of sampled points per strand. For each cluster, we select the strand closest to the cluster center as a salient strand and uniformly resample it into a fixed number $N_\text{salient}$ of points. The union of all sampled points from salient strands forms the salient point set  $\mathcal{P}_\text{salient}$, for which the corresponding binary indicators are set to 1.

\paragraph{3. Uniformly sampled point cloud $\mathcal{P}_\text{uniform}$}
We construct a uniform grid over a bounding box enclosing the hairstyle, with a cell size of $0.2^3 \mathrm{mm}^3$. A single global bounding box is employed to accommodate all hairstyles in the dataset. The center of each grid cell is treated as a candidate point sample. For each candidate $\ss$, we assign an orientation given by the tangent direction of its nearest point in $\mathcal{P}_\text{dense}$. A candidate is discarded if its distance from the nearest point in $\mathcal{P}_\text{dense}$ exceeds a threshold $\epsilon_d$. The remaining samples form the uniformly sampled point cloud $\mathcal{P}^*_{\text{uniform}}$ for hair. In addition, we uniformly sample points within the bust volume to obtain the bust point set $\mathcal{P}_\text{bust}$. We then combine the two sets to form the final uniformly sampled input $\mathcal{P}_\text{uniform} = \mathcal{P}_\text{bust} \bigcup \mathcal{P}^*_\text{uniform}$. Each sample is associated with a binary classification indicator specifying whether it belongs to the bust or to the hair.

\subsection{Training and Inference}

\paragraph{Query point cloud $\mathcal{P}_\text{query}$ for training}
Due to the highly imbalanced distribution of orientation directions in hair regions, uniformly random sampling tends to over-represent dominant directions, thereby introducing bias during training. To address this issue, we propose an \textit{even-sampling strategy} in orientation space. Specifically, we generate $N^\text{dir}$ uniformly distributed unit-direction anchors using a Fibonacci-sphere construction and assign each orientation vector to its nearest anchor. This partitions the orientation space into $N^\text{dir}$ bins, from which samples are drawn evenly. For bust samples, we assign a zero orientation vector $[0,0,0]$. In addition, we apply a spatial importance sampling strategy based on the signed distance to the surface mesh, following prior work~\cite{park2019deepsdf, wang2021neus}.

\paragraph{Training loss}
To train DOAE, we optimize the network parameter set $\psi$ using the following loss function:
\begin{equation}
L_\text{train}(\psi) = \lambda_\mathcal{D}||\hat{\mathcal{D}}-\mathcal{D}||_2^2 + \lambda_\mathcal{O}||\hat{\mathcal{O}}-\mathcal{O}||_2^2 + \lambda_\mathcal{C}{L}_{CE}(\hat{\mathcal{C}}, \mathcal{C}) + \lambda_{KL}{L}_{KL} \; , \notag
\end{equation}
where ${L}_{CE}$ denotes the cross-entropy loss and ${L}_{KL}$ is the KL regularization term used for training VAE~\cite{kingma2013auto}. The ground-truth SDF values $\mathcal{D}$ are computed from the surface mesh. The ground-truth labels of orientation $\mathcal{O}$ and binary indicator $\mathcal{C}$ are assigned from the nearest points in the dense input set $\mathcal{P}_\text{dense}$.

\paragraph{Run-time inference}
As illustrated in \autoref{fig:pipeline}, our method optimizes the latent code $\zz$ at run time using the reconstructed mesh $\mathcal{M}_\text{input}$ and the input image $I$. The optimization is guided by three supervisory signals: SDF values $\mathcal{D}$ computed from randomly sampled 3D points, reprojected orientation cues $\mathcal{O}_\text{proj}$, and screen-space hair segmentation labels $\mathcal{C}_\text{proj}$. We formulate the objective as
\begin{equation} \label{eq:optim}
L_\text{opt}(z) = \lambda_\mathcal{D}||\hat{\mathcal{D}}-\mathcal{D}||_2^2 + \lambda_\mathcal{O}||\hat{\mathcal{O}}-\mathcal{O}_\text{proj}||_2^2 + \lambda_\mathcal{C}L_{CE}(\hat{\mathcal{C}}, \mathcal{C}_\text{proj}).
\end{equation}
where $\zz$ is initialized with the mean latent code of the training set to accelerate convergence.
After optimization, the converged latent code $\zz'$ is passed through the fixed decoder to infer orientation and segmentation fields. We uniformly sample a dense 3D grid (e.g., $512^3$ resolution) and evaluate the decoder to obtain raw predictions $\hat{\mathcal{O}}$ and $\hat{\mathcal{C}}$ (see~\autoref{eq:decoder}). The final hair orientation field $\tilde{\mathcal{O}}$ is extracted via
\begin{equation}
\tilde{\mathcal{O}} = \hat{\mathcal{O}} \odot \texttt{argmax}(\hat{\mathcal{C}}) \odot \Omega,
\end{equation}
where $\odot$ denotes element-wise multiplication, $\texttt{argmax}(\hat{\mathcal{C}})$ yields a binary hair-region mask, and $\Omega$ is the interior occupancy mask of $\mathcal{M}_\text{input}$ computed from its SDF. 

\paragraph{Input labeling}
To compute the reprojected orientation $\mathcal{O}_\text{proj}$ and segmentation $\mathcal{C}_\text{proj}$, we first estimate the screen-space hair orientation $\mathcal{O}_\text{screen}$ and segmentation $\mathcal{C}_\text{screen}$ using HairStep~\cite{zheng2023hairstep}. We then rasterize the input mesh $\mathcal{M}_\text{input}$ onto the image plane $I$, align it using facial landmarks, and record the pixel-to-face correspondences during rasterization. This correspondence enables back-projection of the 2D predictions into 3D space, yielding $\mathcal{O}_\text{proj}$ and $\mathcal{C}_\text{proj}$. In addition to image-based cues, users may optionally provide sparse directional strokes directly on the 3D surface to specify hair orientations. These user-defined directions are treated as ground-truth constraints during optimization. Further implementation details are provided in the appendix.

\subsection{Surface-guided Orientation Correction} \label{sec:correction}

To prevent generated hair strands from penetrating the underlying surface, we introduce a post-processing step that corrects the predicted orientation field using surface normals from the input mesh $\mathcal{M}_\text{input}$. Since hair growth directions should ideally lie within the tangent plane of $\mathcal{M}_\text{input}$, we suppress the normal component of the orientation field from the surface. Specifically, given the predicted orientation field $\tilde{\mathcal{O}}$, we uniformly sample it into a point set $\mathcal{S}=\{\ss_i : [\xx_i,\oo_i,c_i]\}$. For each sample $\ss_i$, we identify its closest point $\hat\xx_i$ on $\mathcal{M}_\text{input}$. If their distance, $d(\hat\xx_i, \xx_i)$, is below the threshold $\epsilon_d$, we project the orientation $\oo$ onto the tangent plane of $\mathcal{M}_\text{input}$ by removing its component along the surface normal. To ensure a smooth transition away from the surface, the projected orientation is blended with the original orientation using a distance-weighted factor. The resulting point set $\mathcal{S}^*=\{\ss_i^* : [\xx_i,\oo_i^*,c_i]\}$ is then used to construct a corrected orientation field $\mathcal{O}^*$ for hair extraction. The pseudocode is presented in Algorithm~\ref{alg:normoptim}. Here, $\alpha$ serves as a scaling factor to constrain the blend weight $w$ within the range $[0.1, 1]$, and $\epsilon_d$ represents a distance threshold set to $16\text{mm}$ in our experiments.

\RestyleAlgo{ruled}
\begin{algorithm} [ht!]
\caption{Normal-based Strand Correction}\label{alg:normoptim}
\DontPrintSemicolon
\newcommand\mycommfont[1]{\footnotesize\ttfamily\textcolor[RGB]{0 128 64}{#1}}
\SetCommentSty{mycommfont} 

\For{$\ss_i=[\xx_i,\oo_i,c_i] \in \mathcal{S}$}{
    find its closest point on $\mathcal{M}_\text{input}$ with position $\hat{\xx}$ and normal $\hat{\nn}$ \\
    \uIf{$d(\hat{\xx}, \xx_i) > \epsilon_d$}{ 
        $\oo^* = \oo$
    } \uElse {
        \tcp{project $\oo$ onto the surface}
        $\oo_\text{proj} = {(\oo-\oo\cdot\hat{\nn})}\;/\;{|\oo-\oo\cdot\hat{\nn}|}$ \\
        $w = {1.0}\;/\;{\bigl(1.0 + \alpha\; d(\hat{\pp}, \pp_i)\bigr)}$ \tcp{Blend weight $w$}
        $\oo^* = w\;\oo_\text{proj} + (1 - w)\;\oo$ \\
        $\oo^* = {\oo^*}\;/\;{|\oo^*|}$ \\
        $\ss_i^*=[\xx_i,\oo^*_i,c_i]$ \\
    }
}
\end{algorithm}

\subsection{Strands Extraction}
\label{sec:strandExtract}

CT2Hair~\cite{shen2023CT2Hair} introduces a hair extraction pipeline based on an orientation field; however, the resulting strands do not necessarily yield an even distribution of hair roots over the scalp. In contrast, methods that grow hair directly from the scalp via advection in the orientation field~\cite{chai2013dynamic, shen2020deepsketchhair, zheng2023hairstep, zhou2024groomcap} often suffer from missing regions and visible gaps in the generated strand set. To leverage the strengths of both paradigms while mitigating their respective limitations, we adopt a hybrid extraction strategy.

Specifically, we first discretize the corrected orientation field $\mathcal{O}^*$ into a dense point cloud and follow the CT2Hair pipeline by applying mean-shift filtering~\cite{nam2019strand}, segment generation, and guided growing to produce a set of guide strands $\mathcal{G}$. In addition, since $\mathcal{O}^*$ explicitly encodes strand directionality, we further densify the hairstyle by tracing additional strands $\mathcal{E}$ from scalp vertices not covered by the guide strands, using forward Newton advection. The final strand-based hair model is obtained by combining both sets, $\mathcal{H}_\text{extract}=\mathcal{G}\bigcup\mathcal{E}$.

\subsection{Orientation Field Refinement}
\label{sec:optim}

Proper initialization of the latent code is crucial for optimization-based latent inversion~\cite{alaluf2021restyle}, especially for inputs that lie outside the training distribution. To address this challenge, we introduce an iterative refinement strategy in which the strands extracted in the first pass are fed back into the encoder to regress a refined latent code for a second optimization stage. Specifically, we use the data synthesis pipeline described in~\autoref{sec:data} to convert the initially extracted strands to the point-cloud format for the encoder input. \Rv{At test time, both $\mathcal{P}_\text{uniform}$ and $\mathcal{P}_\text{salient}$ are generated on the fly from the first-pass strands using the same sampling and $k$-means clustering procedure as during training. As a result, the refinement stage is fully self-contained and does not require ground-truth strands.} The encoder then predicts a refined latent code $\zz^*$, which is used to initialize the subsequent optimization. Starting from $\zz^*$, we minimize the objective function defined in \autoref{eq:optim} to obtain the final result. Empirically, we find that one iteration of this refinement loop is sufficient to achieve convergence since the output of our first pass is already good enough.

%% file: sections/3.exp_arxiv.tex
\section{Experiments}\label{sec:experiments}
In this section, we evaluate HairLRM performance through both quantitative analysis on synthetic datasets and qualitative comparisons on in-the-wild images. We perform comprehensive ablation studies to validate the effectiveness of individual components within our framework.
Furthermore, we compare our approach against state-of-the-art hair modeling methods, including NeuralHDHair~\cite{wu2022neuralhdhair}, HairStep~\cite{zheng2023hairstep}, and DiffLocks~\cite{rosu2025difflocks}, to demonstrate its superiority. A gallery of results is also presented to illustrate the versatility of our method.

We adopt Hunyuan3D~\cite{lai2025hunyuan3d} as the backbone Large Reconstruction Model (LRM) for all experiments. \Rv{Our pipeline is backbone-agnostic since DOAE only consumes the mesh $\mathcal{M}$.} For 2D supervision, we employ HairStep~\cite{zheng2023hairstep} to predict orientation maps and segmentation masks. During inference, we first optimize the orientation field using our pre-trained model for 300 iterations. Subsequently, we perform the proposed orientation field refinement for up to 300 additional iterations, with early stopping applied upon convergence. Please refer to the appendix for further implementation details.

\begin{figure}[t!]
    \centering
    \includegraphics[width=\linewidth]{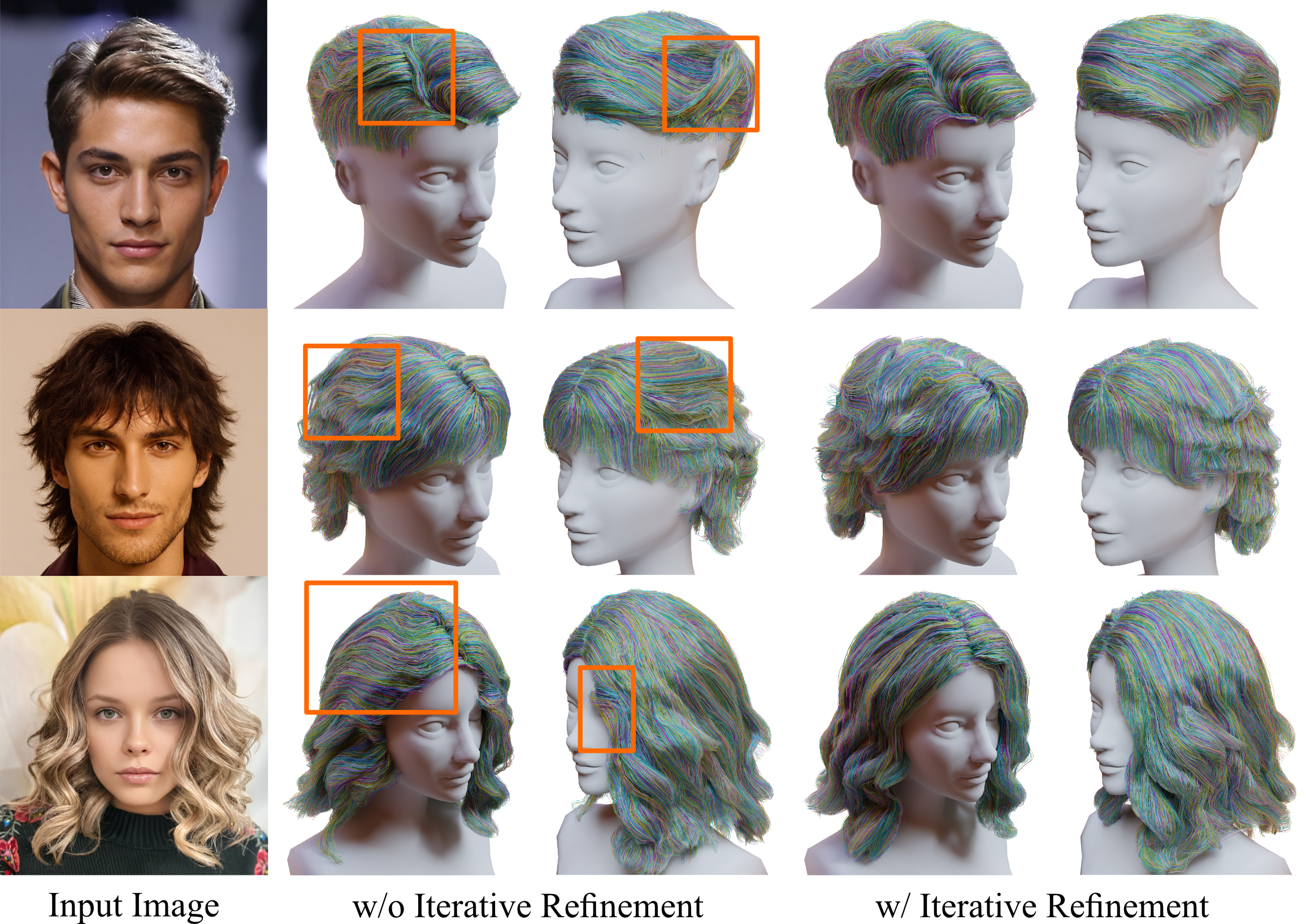}
    \caption{Ablation study on iterative refinement. 
            Middle: Without refinement, the optimization is prone to getting trapped in local minima, leading to unnatural discontinuities and chaotic strands. 
            Right: Our refinement strategy effectively removes these artifacts, yielding coherent, continuous hair strands.}
    \label{fig:AbIterOptim}
    \Description{}
    \vspace{-0.3cm}
\end{figure}

\paragraph{DOAE network}
We first evaluate the effectiveness of the proposed DOAE network by comparing it with a generic backbone, Point Transformer V3 (PtXformerV3)~\cite{wu2024point}. Both models are trained from scratch under identical configurations and evaluated in the Perm test set~\cite {he2025perm}. Using explicit hairstyle annotations provided by the dataset, we report performance separately for curly and straight hair. As shown in \autoref{tab:EvenDiff}, DOAE consistently outperforms the baseline (PtXformerV3 trained without even sampling) on all evaluation metrics. In particular, DOAE achieves an approximately 50\% reduction in MSE, highlighting its superior ability to model complex hair orientation fields.

\paragraph{Orientation field refinement}
Finally, we evaluate the proposed iterative refinement strategy described in \autoref{sec:optim}. To ensure a fair comparison, we configure the baseline (without refinement) to directly optimize the latent code for 600 iterations, matching the total computational budget of our full pipeline. As shown in \autoref{fig:AbIterOptim}, simply increasing the number of optimization steps without refinement yields suboptimal results. The baseline often gets stuck in local minima, leading to unnatural discontinuities and chaotic strand structures that deviate from the intended global hairstyle (e.g., the disordered short hair in Row 1 and the inconsistent curls in Row 2). In contrast, our refinement strategy effectively escapes these local optima, producing coherent strands with smooth connectivity that more faithfully match the input image.

\begin{figure}[t!]
    \centering
    \includegraphics[width=\linewidth]{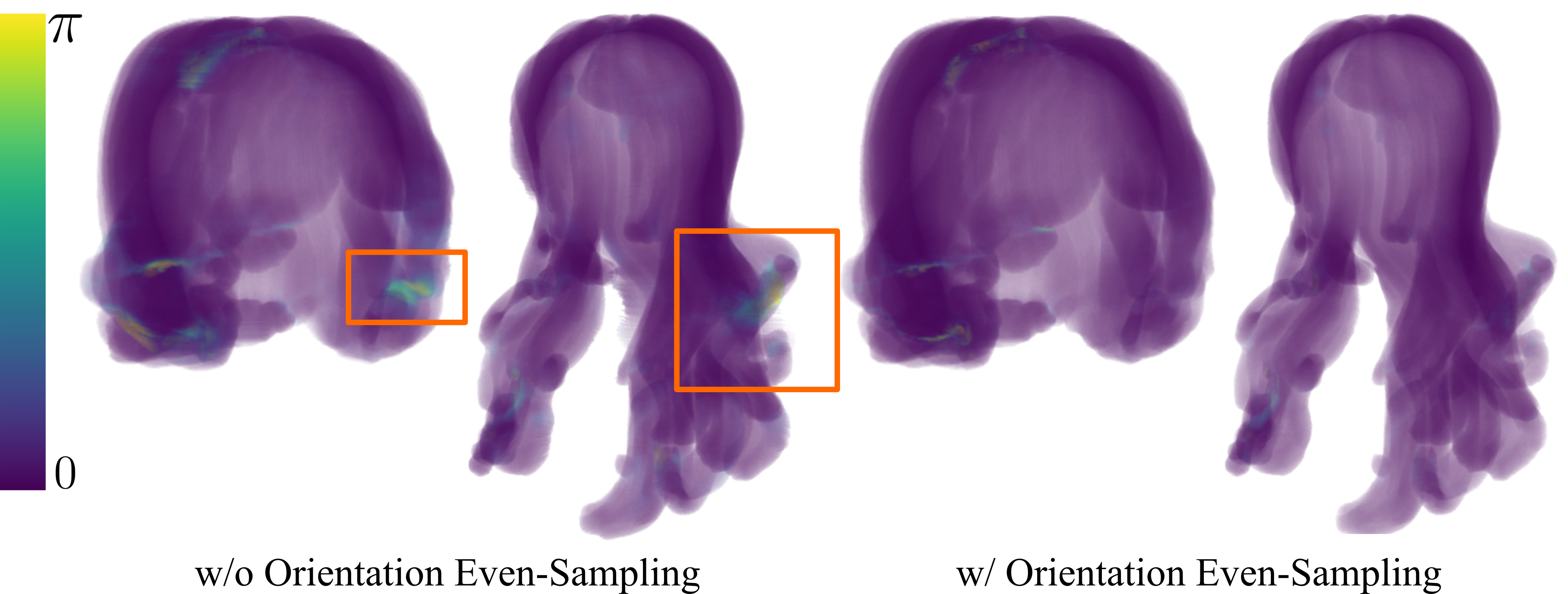}
    \caption{Error map visual comparison of our orientation even-sampling strategy in the training stage. The results using even-sampling show lower angle error values in the non-straight region.}
    \label{fig:AbEvenly}
    \Description{}
    \vspace{-0.3cm}
\end{figure}

\paragraph{Orientation evenly sampling}
We further analyze the impact of our even-sampling strategy through both quantitative and qualitative evaluations. \autoref{tab:EvenDiff} shows that even sampling leads to a significant performance gain, particularly for curly hairstyles, where it reduces MAE by 4.5$^\circ$. Qualitatively, \autoref{fig:AbEvenly} presents per-point error maps, revealing a substantial reduction in errors in high-curvature regions (i.e., non-vertical and non-straight areas). These results validate our design motivation that even sampling effectively alleviates data imbalance in geometrically complex regions.

\begin{figure}[ht!]
    \centering
    \includegraphics[width=\linewidth]{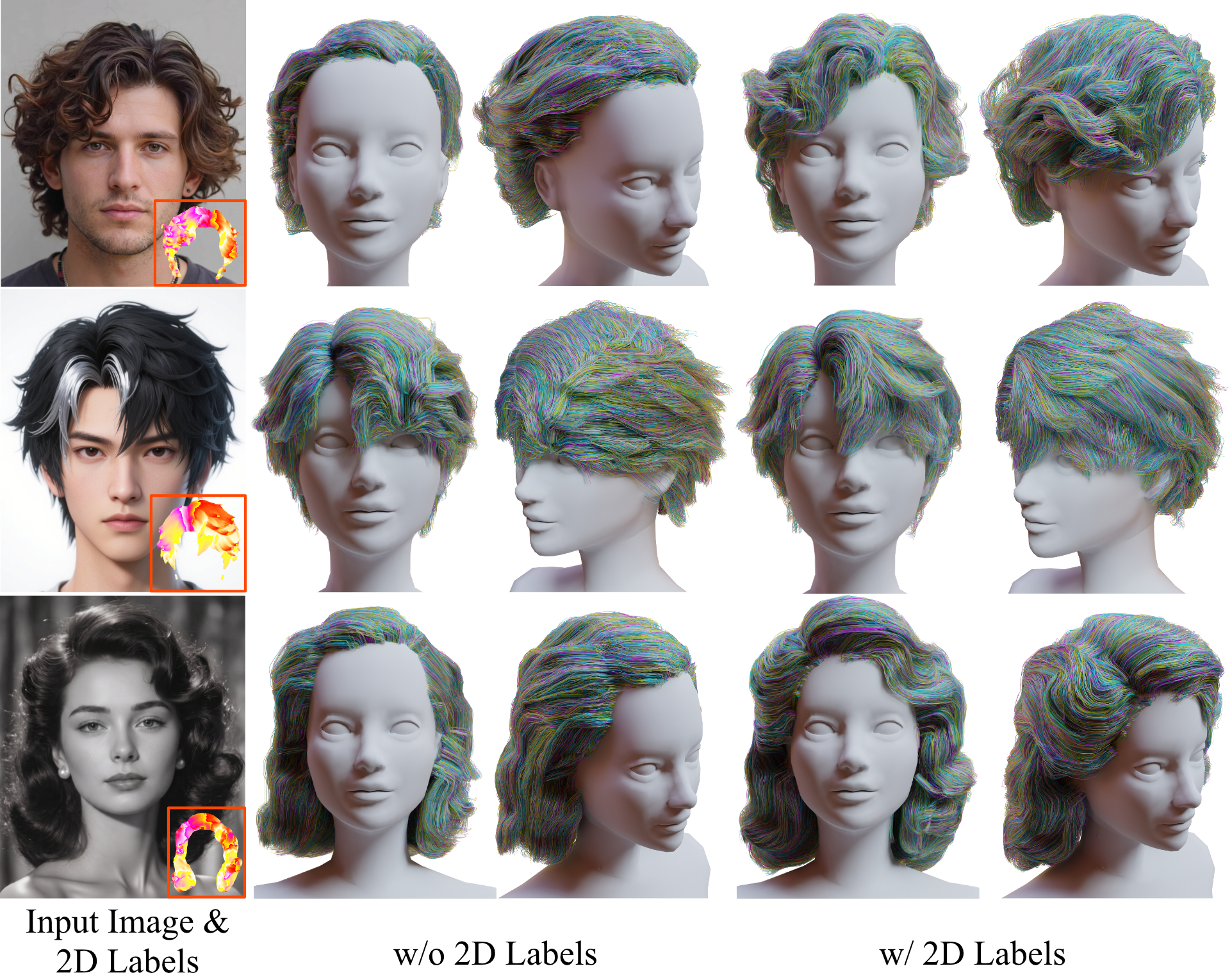}
    \caption{
        Ablation study on 2D labels (i.e., 2D orientations and segmentation in the screen space).
        We evaluate the effectiveness of using 2D orientation and segmentation maps (visualized in the insets of the first column) as optimization targets. 
        Middle: Without 2D labels, the reconstructed strands often deviate from the input image, especially for out-of-domain styles (2nd row) or complex curls (3rd row). 
        Right: With 2D labels enabled, our method generates strands that faithfully align with the input appearance and flow.
    }
    \label{fig:Ab2D}
    \Description{}
\end{figure}

\paragraph{Automatic 2D labels}
We evaluate the necessity of these 2D cues (orientation and segmentation maps generated by HairStep~\cite{zheng2023hairstep}) through an ablation study in which the 2D supervision terms are disabled during optimization (i.e., setting $\lambda_\mathcal{O} = 0$ and $\lambda_\mathcal{C} = 0$), while keeping all other settings unchanged, as shown in \autoref{fig:Ab2D}. Without 2D supervision, the method can recover a coarse global hair shape; however, the reconstructed strands often exhibit noticeable inconsistencies with the input image. In contrast, incorporating 2D labels imposes strong screen-space constraints, yielding strand orientations that more faithfully align with the source appearance. Two representative cases highlight the benefits of this design. Generalization (Row 2): The cartoon hairstyle lies outside the training distribution. Without 2D guidance, the model fails to infer coherent orientations, resulting in disordered strands. The 2D orientation map provides explicit directional cues that enable the generation of consistent, stylized strands. Complex Geometry (Row 3): For a challenging curly hairstyle, the baseline without 2D labels produces visually plausible but incorrect strand directions. The addition of 2D supervision effectively corrects these errors, yielding reconstructions that better match the intricate curl patterns observed in the input image.

\begin{figure}[ht!]
    \centering
    \includegraphics[width=\linewidth]{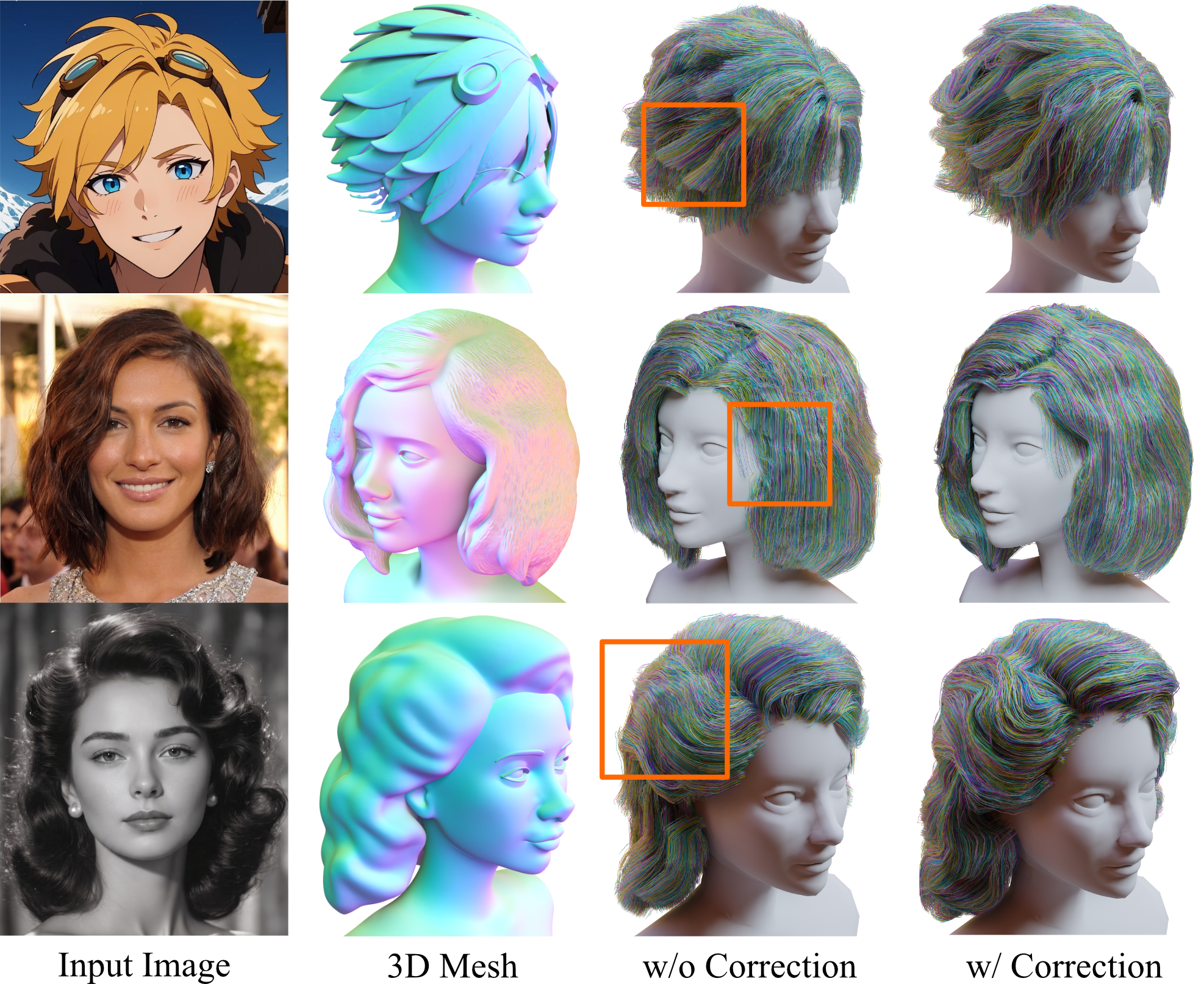}
    \caption{Ablation study on surface-guided orientation correction.
             We compare the results before and after applying the correction step. 
             As highlighted, the correction significantly improves the alignment between hair strands and the underlying mesh surface, thereby improving surface adherence and enhancing strand continuity.}
    \label{fig:AbNormOptim}
    \Description{}
\end{figure}

\paragraph{Surface-guided orientation correction}
We evaluate the orientation correction module, which enforces alignment between the predicted orientation field and the surface normals of the underlying mesh. As illustrated in \autoref{fig:AbNormOptim}, this correction leads to notable improvements in geometric fidelity across several aspects. Clumping Consistency (Row 1): The correction restores coherent clumping patterns that better reflect the geometric features of the input mesh. Surface Adherence (Row 2): Corrected strands adhere more closely to the mesh surface, particularly in lateral regions, effectively eliminating unnatural gaps. Strand Continuity (Row 3): The operation improves strand continuity, yielding smoother, more natural flow compared to the uncorrected results.

\Rv{\paragraph{Dual point-cloud design}
To isolate the contribution of our dual point-cloud formulation, we ablate each stream while keeping the rest of the pipeline fixed: \textit{Uniform-only} removes the salient cross-attention branch, and \textit{Salient-only} removes the uniform branch. As~\autoref{tab:EvenDiff} shows, both variants substantially degrade over the full model, reaching MAE $12.78^\circ$ and $13.37^\circ$ (44.2\% and 46.7\% relative increase, respectively). Neither stream alone suffices; the uniform stream anchors the global hair volume, while the salient stream captures high-frequency directional structure, and only their dual-attention fusion resolves both.}

\newcommand{\sd}[1]{\ensuremath{\,\text{\tiny$\pm\,#1$}}}

\begin{table}[ht!]
\centering
\caption{Quantitative comparisons of DOAE variants on the Perm~\cite{he2025perm} test set. \textit{ES} denotes our even-sampling strategy during training. \Rv{\textit{Uni-only} and \textit{Sal-only} ablate the dual point-cloud design by removing one of the two complementary point clouds (\textit{Uniform-only} and \textit{Salient-only}, respectively); our full model fuses both via dual cross-attention.} MSE, MAE, and $\overline{\text{MAE}}$ denote Mean Squared Error, Mean Angular Error, and Median Angular Error, respectively. Results are reported as mean $\pm$ std across evaluation samples.}
\small
\begin{tabular}{l|c|ccc}
\toprule
\textbf{Methods} & \textbf{Metric} & \textbf{Curly} & \textbf{Straight} & \textbf{TOTAL} \\
\midrule
PtXformerV3 & MSE
& $0.052\sd{0.029}$ & $0.031\sd{0.013}$ & $0.040\sd{0.023}$ \\

\Rv{DOAE Uni-only} & \Rv{MSE}
& \Rv{$0.041\sd{0.020}$} & \Rv{$0.021\sd{0.009}$} & \Rv{$0.029\sd{0.017}$} \\

\Rv{DOAE Sal-only} & \Rv{MSE}
& \Rv{$0.045\sd{0.019}$} & \Rv{$0.023\sd{0.013}$} & \Rv{$0.032\sd{0.019}$} \\

\rowcolor{blue!8}
DOAE w/o \textit{ES} & MSE
& $0.030\sd{0.012}$ & $0.015\sd{0.006}$ & $0.021\sd{0.012}$ \\

\rowcolor{green!8}
\Rv{DOAE (Full)} & MSE
& $0.015\sd{0.007}$ & $0.008\sd{0.003}$ & $0.011\sd{0.007}$ \\

\midrule
PtXformerV3 & MAE
& $17.00^\circ\sd{4.60^\circ}$ & $13.29^\circ\sd{2.79^\circ}$ & $14.82^\circ\sd{4.06^\circ}$ \\

\Rv{DOAE Uni-only} & \Rv{MAE}
& \Rv{$15.07^\circ\sd{3.42^\circ}$} & \Rv{$11.25^\circ\sd{2.16^\circ}$} & \Rv{$12.78^\circ\sd{3.30^\circ}$} \\

\Rv{DOAE Sal-only} & \Rv{MAE}
& \Rv{$16.04^\circ\sd{3.47^\circ}$} & \Rv{$11.59^\circ\sd{2.71^\circ}$} & \Rv{$13.37^\circ\sd{3.73^\circ}$} \\

\rowcolor{blue!8}
DOAE w/o \textit{ES} & MAE
& $12.56^\circ\sd{2.57^\circ}$ & $9.46^\circ\sd{1.56^\circ}$ & $10.70^\circ\sd{2.73^\circ}$ \\

\rowcolor{green!8}
\Rv{DOAE (Full)} & MAE
& $8.19^\circ\sd{1.62^\circ}$ & $6.42^\circ\sd{1.10^\circ}$ & $7.13^\circ\sd{1.80^\circ}$ \\

\midrule
PtXformerV3 & $\overline{\text{MAE}}$
& $12.36^\circ\sd{2.68^\circ}$ & $10.27^\circ\sd{2.32^\circ}$ & $11.13^\circ\sd{2.67^\circ}$ \\

\Rv{DOAE Uni-only} & $\overline{\text{MAE}}$
& \Rv{$11.31^\circ\sd{2.34^\circ}$} & \Rv{$9.04^\circ\sd{1.81^\circ}$} & \Rv{$9.95^\circ\sd{2.32^\circ}$} \\

\Rv{DOAE Sal-only} & $\overline{\text{MAE}}$
& \Rv{$12.05^\circ\sd{2.75^\circ}$} & \Rv{$9.25^\circ\sd{2.09^\circ}$} & \Rv{$10.37^\circ\sd{2.74^\circ}$} \\

\rowcolor{blue!8}
DOAE w/o \textit{ES} & $\overline{\text{MAE}}$
& $9.42^\circ\sd{1.83^\circ}$ & $7.57^\circ\sd{1.26^\circ}$ & $8.31^\circ\sd{2.00^\circ}$ \\

\rowcolor{green!8}
\Rv{DOAE (Full)} & $\overline{\text{MAE}}$
& $5.85^\circ\sd{1.12^\circ}$ & $4.92^\circ\sd{0.85^\circ}$ & $5.29^\circ\sd{1.28^\circ}$ \\
\bottomrule
\end{tabular}
\label{tab:EvenDiff}
\end{table}

\paragraph{Comparison with SOTAs}

We compare our method against four state-of-the-art single-view hair reconstruction approaches.%
For relatively simple straight hairstyles (e.g., \autoref{fig:Comp}a), NeuralHDHair~\cite{wu2022neuralhdhair} produces plausible strand reconstructions. However, its performance degrades on more complex geometries. In particular, for curly hairstyles (\autoref{fig:Comp}b) and non-standard topologies such as ponytails (\autoref{fig:Comp}e,f), the method fails to recover coherent strand structures. In contrast, our approach generalizes robustly across a wide range of hairstyles.
Compared with HairStep~\cite{zheng2023hairstep}, which predicts accurate 2D orientation fields from images, the feed-forward reconstruction network often struggles to recover the correct global 3D structure (\autoref{fig:Comp}a,e,f). Our method instead incorporates these 2D orientation cues as supervision within an optimization-based framework, rather than directly regressing 3D geometry. As a result, our reconstructions (\autoref{fig:Comp}b,c,d) exhibit strand flows that are more consistent with the input images.
DiffLocks~\cite{rosu2025difflocks} is a recent diffusion-based method that predicts per-strand features directly from a single image. Although this generative formulation produces naturally distributed hair, it frequently hallucinates geometric details that deviate from input, leading to reduced fidelity (\autoref{fig:Comp}a,b). Moreover, DiffLocks often fails to capture fine-scale local structures, such as large wavy patterns (\autoref{fig:Comp}c) or layered curls (\autoref{fig:Comp}f). In contrast, our method achieves a better balance between capturing high-frequency details and maintaining strict consistency with the input image.
\Rv{Im2Haircut~\cite{sklyarova2025im2haircut} is a concurrent work that directly regresses strands from a single image via a learned hair prior. Although it recovers reasonable global shape on standard portraits, its reliance on image observations alone tends to smooth out high-frequency directional detail on complex cases such as tight curls and stylized inputs (\autoref{fig:Comp}b,d), whereas our LRM-based geometric scaffold provides an explicit 3D anchor that better preserves such structures.}
Finally, our approach is the first to robustly reconstruct ponytail and twin-ponytail hairstyles from single images, as demonstrated in \autoref{fig:teaser} and \autoref{fig:Comp}e,f, highlighting the effectiveness of HairLRM for challenging hair reconstruction scenarios.
\Rv{\autoref{tab:EvenDiff} shows a quantitative comparison of our HairLRM against the above methods. Our method achieves the best IoU ($0.84$) and lowest orientation error ($18.8^\circ$), clearly outperforming Im2Haircut ($0.76$/$22.0^\circ$) and HairStep ($0.62$/$27.5^\circ$). These gains are consistent with the qualitative observations above, particularly on ponytails, cartoons, and tight curls.}

\begin{table}[ht!]
\centering
\caption{\Rv{Quantitative comparison with SOTA single-view strand reconstruction methods. Following Im2Haircut~\cite{sklyarova2025im2haircut}, we render each reconstruction under the input view and compare against the input: IoU on the hair silhouette (shape) and mean angular error (Ori-Err) against the HairStep~\cite{zheng2023hairstep} 2D orientation of the input (flow). All 3D results are aligned to a common bust.}}
\small
\begin{tabular}{l|cc}
\toprule
\textbf{Method} & \textbf{IoU $\uparrow$} & \textbf{Ori-Err $\downarrow$} \\
\midrule
NeuralHDHair~\cite{wu2022neuralhdhair} & $0.54$ & $33.7^\circ$ \\
HairStep~\cite{zheng2023hairstep} & $0.62$ & $27.5^\circ$ \\
DiffLocks~\cite{rosu2025difflocks} & $0.57$ & $34.6^\circ$ \\
Im2Haircut~\cite{sklyarova2025im2haircut} & $0.76$ & $22.0^\circ$ \\
\rowcolor{green!8}
\textbf{HairLRM (Ours)} & $\mathbf{0.84}$ & $\mathbf{18.8^\circ}$ \\
\bottomrule
\end{tabular}
\label{tab:sota}
\end{table}

\begin{figure}[ht!]
    \centering
    \includegraphics[width=\linewidth]{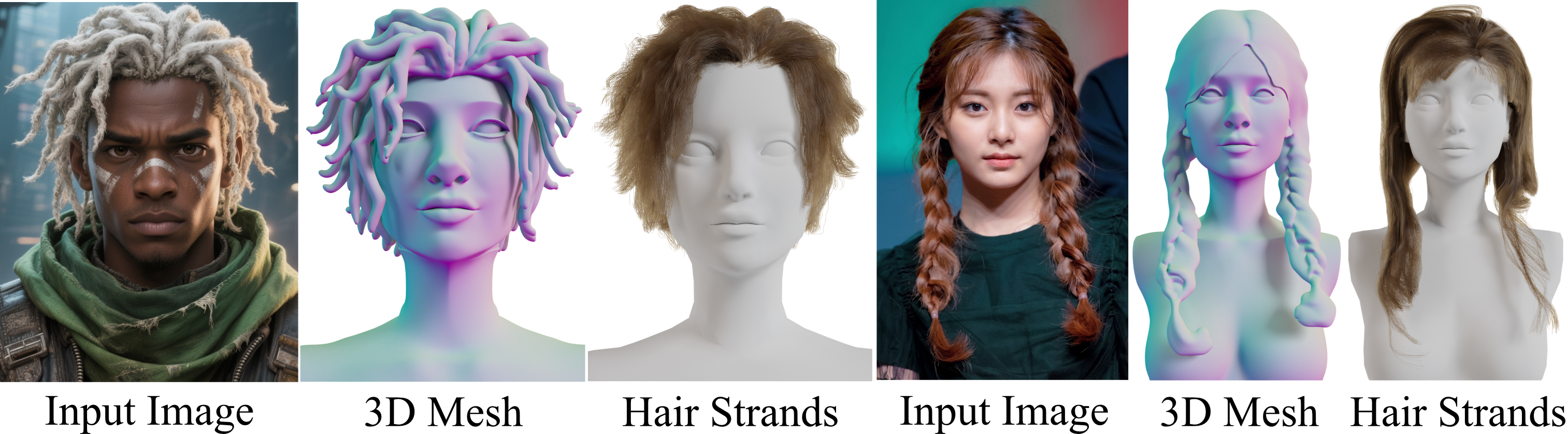}
    \caption{\Rv{Failure cases.}}
    \vspace{-2em}
    \label{fig:fail}
    \Description{}
\end{figure}

\paragraph{Results gallery}
We present a gallery of results in~\autoref{fig:gallery} that covers a wide range of hairstyles, including straight, wavy, curly, and ponytail configurations. These examples demonstrate our method’s ability to generate high-fidelity hair strands with rich geometric detail while faithfully preserving the identity of the input hairstyle.

\Rv{
\paragraph{Failure case}
We observe occasional strand/mesh gaps near hair tips in unconstrained portraits. This missing hair-end artifact is partly due to slight template misalignment, but, more importantly, stems from inaccurate camera intrinsics in in-the-wild data rather than from the sampling scale used during synthesis. Errors in the intrinsics lead to deviations in the reprojected segmentation masks, which in turn manifest as strand/mesh misalignment (\autoref{fig:fail}).}

\Rv{
\paragraph{Runtime performance}
Our optimization-based inference remains the primary runtime bottleneck, taking approximately 15 minutes per portrait on our test device. The cost is dominated by latent optimization ($\sim\!10$ minutes), followed by LRM mesh generation ($\sim\!3$ minutes), hair growing (1--2 minutes), and refinement plus preprocessing ($<35$ seconds). In addition, our method requires substantially less memory at inference time than during training, using about $\sim\!4$~GB VRAM for inference versus $\sim\!80$~GB for training.
}

\paragraph{Applications}
\autoref{fig:cards} illustrates the practicality of our reconstruction for downstream applications. We convert the generated strands into hair cards using the method of~\citet{zheng2025auto}.

%% file: sections/4.conclusion.tex
\section{Conclusion}
We have introduced HairLRM, a novel autoencoder-based framework for high-fidelity reconstruction of hair strands. By leveraging a Large Reconstruction Model (LRM) to generate a coarse 3D mesh, our approach effectively resolves the inherent shape ambiguities present in single-view inputs. Building on this global prior, we have proposed a Dual Orientation AutoEncoder (DOAE) that bridges the representation gap between surface geometry and dense strand-based hair. To further improve geometric fidelity, we have introduced two complementary strategies, surface-guided correction and iterative orientation refinement.

\paragraph{Limitations and future works}
\Rv{Since our pipeline treats the LRM mesh as an explicit scaffold, reconstruction quality is upper-bounded by the mesh itself, and it may also struggle on under-represented styles such as tightly curled braids~\cite{hu2014} or afro-textured hair; broader training data and finer-grained priors are natural next steps. 
We chose an optimization-based framework for two strategic reasons. First, it naturally handles irregular inputs (varying views, sparse user strokes) and interactive editing, both of which are non-trivial for fixed-input feed-forward networks. Second, for faster model generation, distilling the optimized latent manifold into a feed-forward amortized decoder is a promising direction for future acceleration.}

%% file: sections/5.appendix.tex
\section{Implementation Details}
\noindent \textit{Hardware and Training.}
We conduct all experiments on high-end NVIDIA GPUs equipped with 14,592 CUDA cores and 96GB VRAM. For DOAE training, we use 4 GPUs in parallel with a batch size of 32. We employ the Adam optimizer with an initial learning rate of $1e^{-4}$. 
During training, we sample $100\text{K}$ query points per hairstyle at each iteration (following our even-sampling strategy) to compute the reconstruction loss. The entire training process takes approximately 72 hours to converge over $300\text{K}$ iterations.

\noindent \textit{Inference.}
During inference, the optimization process is performed on a single GPU. We maintain the sampling density at $100\text{K}$ query points per iteration. The latent code is optimized for 300 iterations in both the initial optimization (1st round) and the refinement stages. 
The total inference time is approximately 10--15 minutes per subject.

\noindent \textit{Loss weights.}
We use consistent loss weights for both training and runtime optimization: $\lambda_{\mathcal{D}}=10.0$, $\lambda_{\mathcal{O}}=0.02$, and $\lambda_{CE}=0.01$. 
The KL-divergence weight is set to $\lambda_{KL}=0.001$ exclusively during the training phase.

\noindent \textit{Dataset Synthesis Parameters.}
Regarding the dataset synthesis described (Sec.~3.2 in the main paper), we set the number of strand clusters to $N_{\text{cluster}}=256$. 
For the salient point cloud, we sample $N_{\text{salient}}=50\text{K}$ points from the salient strands. For the uniformly sampled point cloud $\mathcal{P}_{\text{uniform}}$, the number of points $N_{\text{uniform}}$ ranges from $500\text{K}$ (for short hair) up to $2\text{M}$ (for long hair).

\section{Runtime Details}

\subsection{Reprojection of 2D features}
\label{sec:reprojection}
This section details the implementation of the 2D-to-3D feature reprojection pipeline described in the main paper. The process consists of three stages: high-resolution 2D inference, rasterization-based correspondence, and differential vector lifting.

\paragraph{2D Inference and Segmentation}
To ensure precise alignment between the 2D observation and the 3D geometry, we employ a rigorous segmentation strategy. We utilize the Segment Anything Model (SAM~\cite{kirillov2023segmentanything}, same as HairStep~\cite{zheng2023hairstep}) to extract binary hair masks from both the input RGB image $\mathcal{C}_\text{input}$ and the rendered view of the mesh $\mathcal{C}_\text{render}$. The final validity mask is computed as the intersection of these two regions: $\mathcal{C}_\text{screen} = \mathcal{C}_\text{input} \cap \mathcal{C}_\text{render}$. This intersection guarantees that feature reprojection is restricted to regions where the 2D visual cues are consistent with the current 3D geometry, effectively filtering out background noise and occluded areas. The orientation map $\mathcal{O}_\text{screen}$ is then predicted from the masked image using HairStep~\cite{zheng2023hairstep}.

\paragraph{Rasterization and Correspondence}
We first utilize the aligned generated 3D mesh $\mathcal{M}$ alongside the blend image $\mathcal{I}$ to derive the camera parameters. 
Note that since we reuse the intrinsic matrix from the template bust rendering, only the extrinsic matrix requires computation via model translation. 
(The alignment process for the generated 3D mesh is detailed in \autoref{subsec:mesh_fitting}.) 
To establish a correspondence between screen space and 3D space, we rasterize a pixel-to-face map $\mathcal{I}_\text{p2f}$. 
This map stores the index of the visible mesh triangle $f_i$ at pixel coordinates $(u_i, v_i)$, implemented using differentiable rendering frameworks such as PyTorch3D~\cite{ravi2020pytorch3d}.

\begin{figure}[t]
\centering
\includegraphics[width=\linewidth]{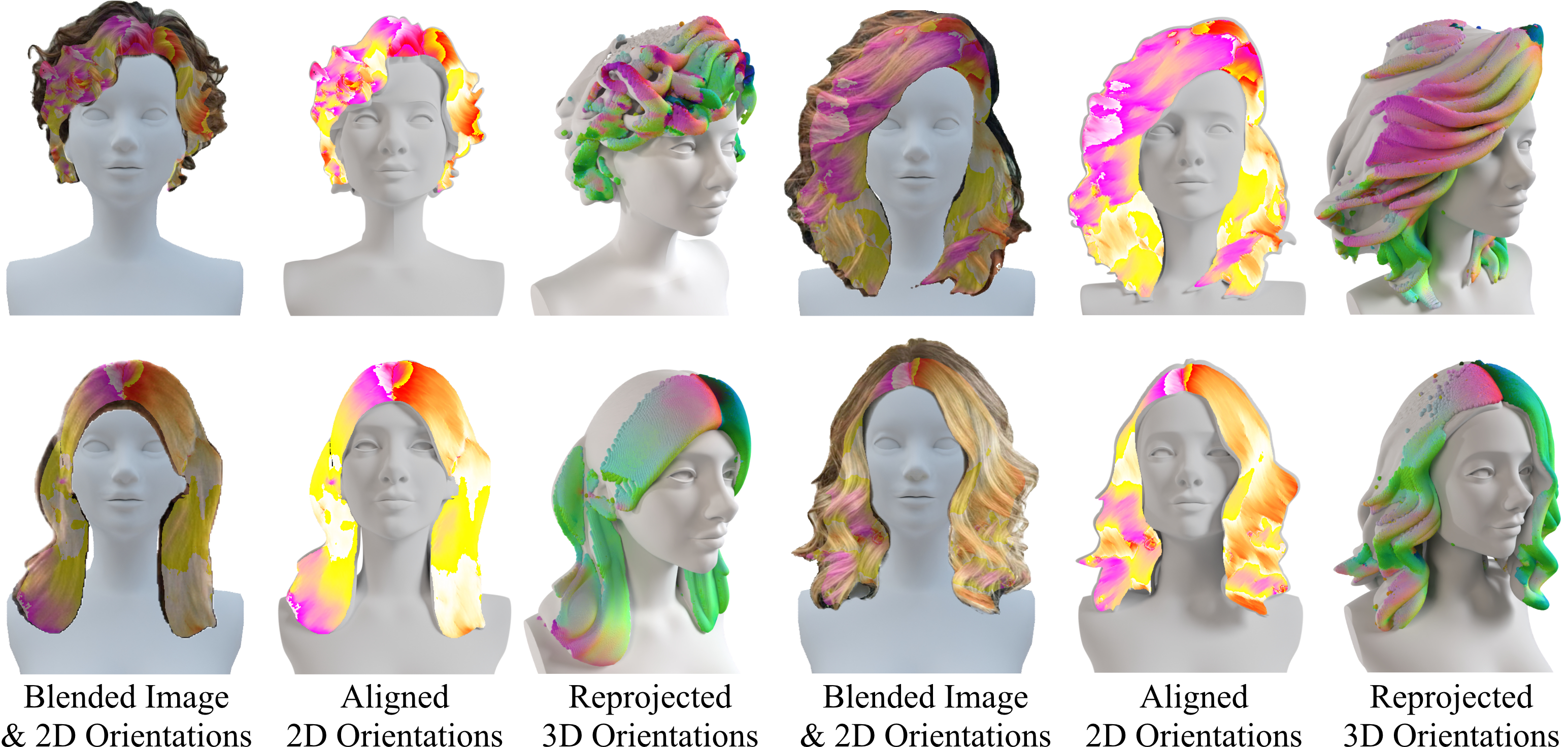}
\caption{Reprojection of 2D features via rasterization-based correspondence and differential vector lifting.}
\vspace{-0.3cm}
\label{fig:reprojection}
\Description{}
\end{figure}

\paragraph{Differential Vector Lifting}
Directly reprojecting a 2D orientation vector is ambiguous due to the loss of depth information. We implement a differential lifting strategy to robustly recover the 3D tangent directions. 
For each sampled pixel $p_i$, we take the 2D orientation vector $\mathbf{o}_i=\mathcal{O}_{\text{screen}}(u_i,v_i)\in\mathbb{R}^2$ and normalize it as $\hat{\mathbf{o}}_i=-\mathbf{o}_i/\|\mathbf{o}_i\|_2$. We then step along this direction to obtain a neighboring pixel:
\begin{equation}
p_{\text{next}}=\big(\lfloor u_i+\delta\,\hat{o}_{u,i}\rfloor,\ \lfloor v_i+\delta\,\hat{o}_{v,i}\rfloor\big).
\end{equation}
Specifically, we set $\delta=2$ pixels.
By querying the \textit{pixel-to-face} mapping $\mathcal{I}_\text{p2f}$, we first locate the underlying mesh triangles $f_i$ and $f_{\text{next}}$ for the sampled pixels. We then extract their corresponding 3D centroids to serve as the spatial coordinates $\mathcal{P}_i$ and $\mathcal{P}_{\text{next}}$.
The 3D reprojected orientation $\mathcal{O}_{\text{proj}}$ is then derived as the normalized difference vector:
\begin{equation}
    \mathcal{O}_{\text{proj}} = \texttt{normalize}(\mathcal{P}_{next} - \mathcal{P}_i)
\end{equation}
As shown in \autoref{fig:reprojection}, our method effectively utilizes the discrete surface manifold to resolve the 3D direction consistent with the 2D orientation.

\subsection{3D Mesh Fitting}
\label{subsec:mesh_fitting}

To ensure that the input scan is rigorously aligned with our canonical template space, we employ a coarse-to-fine registration pipeline. This process is designed to be robust against variations in hairstyle and acquisition noise by prioritizing stable facial and somatic features over potentially variable hair geometry.

\paragraph{3D Anchor Landmark}
Since the hair region often lack semantic correspondence, we firstly recover a set of 3D facial landmarks as the anchor for coarse alignment.
We render the input mesh from multiple viewpoints and utilize MediaPipe~\cite{lugaresi2019mediapipeframeworkbuildingperception} to predict 2D facial landmarks $\{\mathcal{P}^v \}$ in each view $v$. 
Utilizing the known camera intrinsic $K$ and extrinsics $[R|T]$ from the rendering setup, we triangulate these multi-view observations to obtain a set of high-confidence 3D facial landmarks $\mathcal{P}_{\text{face}}$. 
This step effectively lifts the robust 2D facial landmarks into 3D space, providing a reliable anchor for initial alignment.

\paragraph{Coarse Alignment via RANSAC}
We employ RANSAC~\cite{fischler1981ransac} to estimate a coarse transformation $\mathcal{T}_{\text{coarse}}$ between the facial landmarks $\mathcal{P}_{\text{face}}$ and the target canonical landmarks $\mathcal{P}_{\text{tgt}}$.
The transformation $\mathcal{T}_{\text{coarse}}$ is iteratively refined until the inlier ratio exceeds a strict threshold $\tau$. 
This provides a robust initialization that places the mesh in the correct orientation and scale to handle potential outliers caused by detection noise or extreme expressions.

\paragraph{Region-Aware ICP Refinement}
Initialized with $\mathcal{T}_{\text{coarse}}$, we utilize ICP~\cite{besl1992icp} to optimize $\mathcal{T}_{\text{refine}}$ to refine the alignment between the generated mesh and our canonical template space.
Directly applying ICP still suffers from misalignment due to the generated mesh differing from the target one.
To mitigate this, we heuristically isolate the stable body region below the head, ensuring the alignment is anchored to the bust without interference from the variable hair shape. We then combine these body points with the high-confidence facial landmarks to form a robust registration source. The final transformation $\mathcal{T}_{\text{final}} = \mathcal{T}_{\text{refine}} \cdot \mathcal{T}_{\text{coarse}}$ is computed via point-to-point ICP and applied to the full mesh.

\section{Optional Manual Labeling at Runtime}
\label{sec:manual_labeling}

\begin{figure}[t!]
\centering
\includegraphics[width=\linewidth]{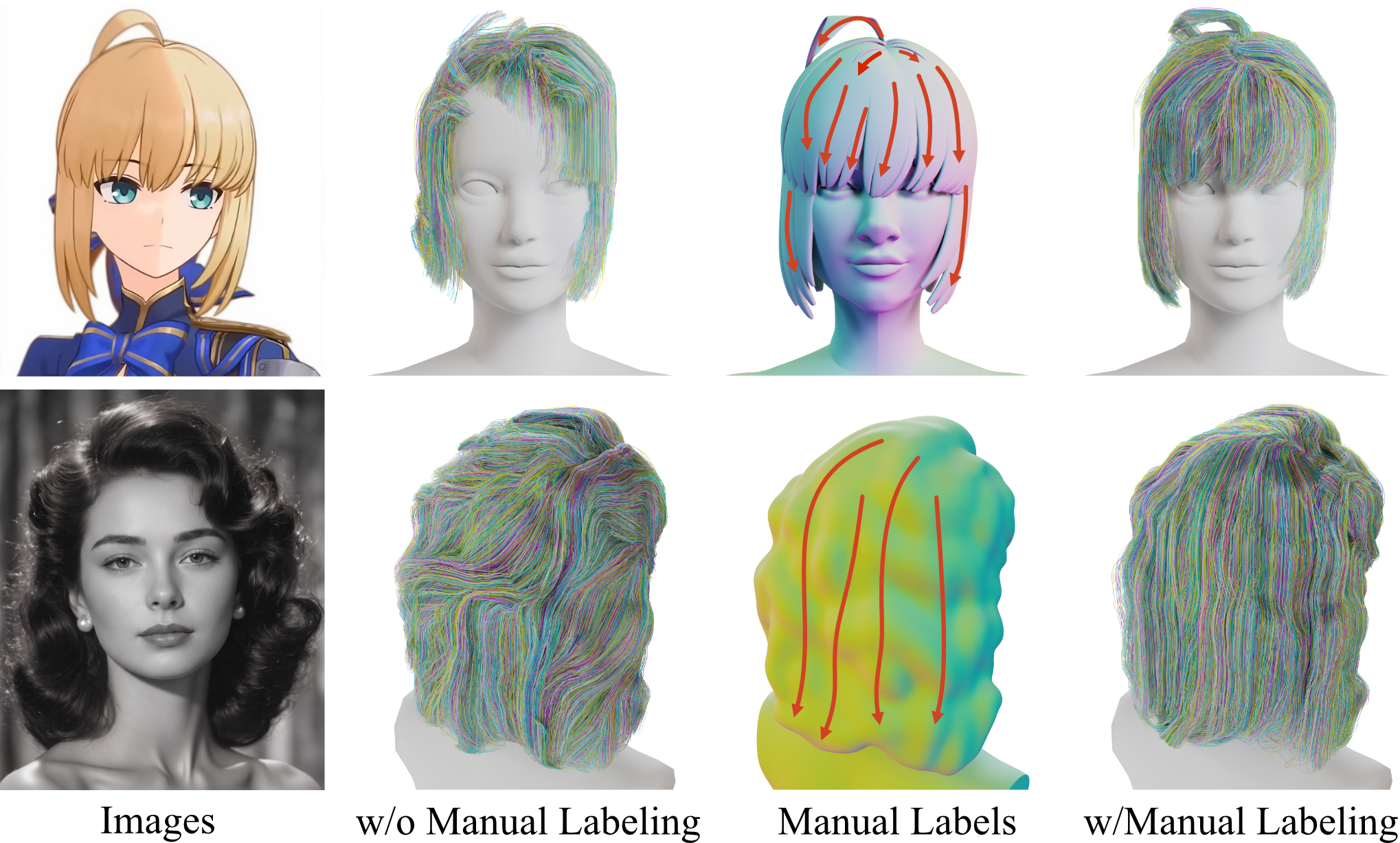}
\caption{
This figure illustrates the impact of optional manual labeling on our results. 
By integrating sparse user strokes, our method effectively recovers missing information inherent to single-view inputs and resolves ambiguities in regions with absent or uncertain 2D orientation predictions, thereby refining local geometric details. 
Notably, as shown in the first row, our method yields plausible results even when the estimated 2D orientation map is replaced by user annotations.
}
\label{fig:manual_label}
\Description{}
\vspace{-0.3cm}
\end{figure}

\begin{figure}[t!]
\centering
\includegraphics[width=\linewidth]{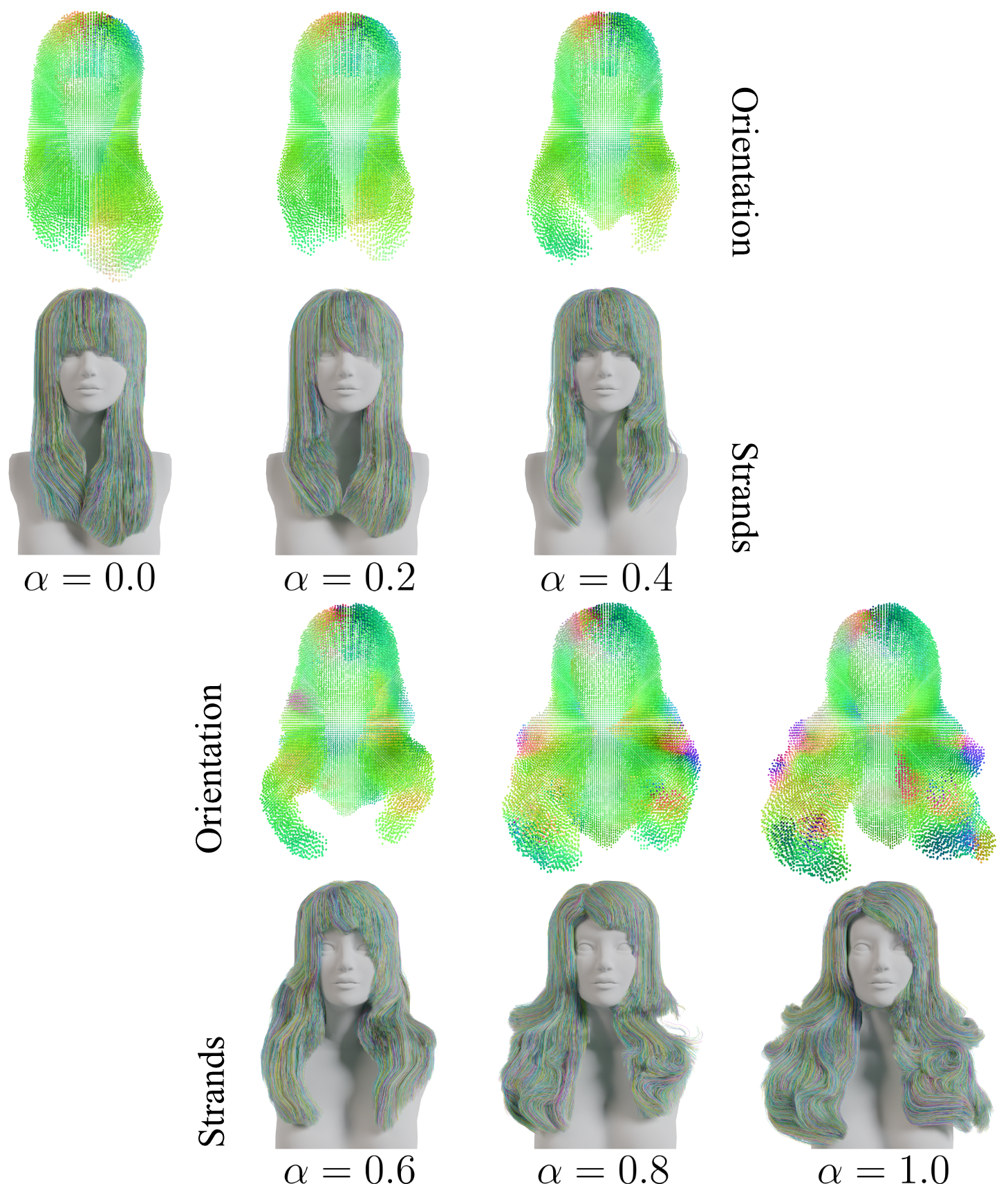}
\vspace{-0.2cm}
\caption{
Latent space interpolation for hairstyle mixing. The interpolation factor $\alpha \in [0,1]$ controls the mixing, where $\alpha=0$ and $\alpha=1$ correspond to two different hairstyles. Intermediate samples visualize smooth interpolation for orientation and strands between the two 
hairstyles.
One can observe the hair tips initially clustered together before gradually separating. Concurrently, the bangs transition from a straight style to a side-swept configuration.
}
\label{fig:latent_interpolation}
\Description{}
\vspace{-0.3cm}
\end{figure}

While the automated pipeline provides dense orientation priors, complex hairstyles often require precise local control. We provide a runtime interface that allows users to introduce sparse, ground-truth constraints via manual strokes.

\paragraph{Interactive Annotation Interface}
Our interactive tool renders the mesh from a user-selected viewpoint using the same rasterization setup described in \autoref{sec:reprojection}. We allow users to add directional labels by simply drawing strokes $\mathcal{S} = \{p_1, p_2, \dots, p_k \}$ in screen space, where each $p_i = (u_i, v_i)$ represents a pixel coordinate (as shown in \autoref{fig:manual_label}). These strokes are intended to mimic the natural flow of hair strands.

\paragraph{Tangent Estimation and 3D Lifting}
For each drawn stroke, we first compute the screen-space tangent vectors $t_{sreen}$ using finite differences between consecutive recorded points: $t_{i} = \text{normalize}(p_{i+1} - p_{i})$. 
We treat these 2D directives as the same 2D orientation labels mentioned in \autoref{sec:reprojection}, and we can calculate the 3D orientation in the same way.

\paragraph{Multi-View Aggregation}
Given the inherent complexity and self-occlusion of hair geometry, a single input view often proves insufficient. 
Consequently, our framework supports multi-view annotation. Users can specify desired 2D orientations from arbitrary viewpoints by drawing strokes. 
The second row of \autoref{fig:manual_label} illustrates the improvement achieved through user annotation from a back view. 
Similar to the reprojected 3D orientations, annotated constraints from all views are aggregated into a unified point cloud $\mathcal{P}_{\text{man}} = \{\mathcal{P}_v\}_{v=1}^{N_{\text{view}}}$. 
Each point in this set encodes the user-defined 3D position and orientation, serving as a ground-truth constraint for optimizing the latent code.

\section{Latent Space Exploration}

To better understand the semantics of our learned hair latent space, we perform latent interpolation for hairstyle mixing. Given two hairstyles $\mathcal{H}^A$ and $\mathcal{H}^B$, we encode them into the decoder’s intermediate latent space, yielding latent codes $\mathbf{z}^{A}$ and $\mathbf{z}^{B}$. We then do the interpolation following:
\begin{equation}
    \mathbf{z}(\alpha) = (1-\alpha)\,\mathbf{z}^{A} + \alpha\,\mathbf{z}^{B},\quad \alpha\in[0,1],
\end{equation}
decode $\mathbf{z}(\alpha)$ to obtain intermediate 3D orientation fields, and then extract hair strands from them.

As shown in \autoref{fig:latent_interpolation}, the results remain plausible across the entire range of $\alpha$. This exploration suggests a well-structured and semantically meaningful latent manifold, enabling controllable hairstyle transitions for applications such as hairstyle editing in future work.